\documentclass{article}

\PassOptionsToPackage{numbers, compress}{natbib}
\usepackage[dblblindworkshop, final]{neurips_2025}
\workshoptitle{Deep Learning for Code in the Agentic Era}

\usepackage[utf8]{inputenc} % allow utf-8 input
\usepackage[T1]{fontenc}    % use 8-bit T1 fonts
\usepackage{hyperref}       % hyperlinks
\usepackage{url}            % simple URL typesetting
\usepackage{booktabs}       % professional-quality tables
\usepackage{amsfonts}       % blackboard math symbols
\usepackage{nicefrac}       % compact symbols for 1/2, etc.
\usepackage{microtype}      % microtypography
\usepackage[dvipsnames]{xcolor}

\title{The Complexity Trap: Simple Observation Masking Is as Efficient as LLM Summarization for Agent Context Management}

\author{%
Tobias Lindenbauer$^{1,2}$ \quad Igor Slinko$^{1}$ \quad Ludwig Felder$^{2}$ \\ 
\textbf{Egor Bogomolov}$^1$ \quad \textbf{Yaroslav Zharov}$^1$\\
$^1$JetBrains Research \\
$^2$School of Computation, Information and Technology, Technical University of Munich \\
\texttt{tobias.lindenbauer@jetbrains.com}
}

\usepackage[nolist,nohyperlinks]{acronym}
\usepackage{subcaption}
\usepackage{amsmath}
\usepackage{booktabs}
\usepackage{hyperref}
\usepackage{cleveref}
\usepackage{graphicx}
\usepackage[nolist,nohyperlinks]{acronym}
\usepackage{xcolor}
\usepackage{pifont}
\usepackage{wrapfig}
\usepackage{multirow}   % To span rows for model names
\usepackage{siunitx}
\usepackage[frozencache]{minted}
\usepackage{tcolorbox}
\definecolor{jb_magenta}{HTML}{FF318C}
\definecolor{jb_magenta_accent}{HTML}{FFCEE4}
\definecolor{jb_violet}{HTML}{6B57FF}

\begin{acronym}
\acro{llm}[LLM]{Large Language Model}
\acro{lrm}[LRM]{Large Reasoning Model}
\acro{slm}[SLM]{Small Language Model}
\acro{cot}[CoT]{Chain-of-Thought}
\acro{se}[SE]{Software Engineering}
\acro{pl}[PL]{Programming Language}
\acro{mcp}[MCP]{Model-Context Protocol}
\acro{qa}[QA]{Question-Answering}
\acro{cua}[CUA]{Computer-Use Agent}
\end{acronym}

\begin{document}

\maketitle

\begin{abstract}
  % Rationale & Problem
  \ac{llm}-based agents solve complex tasks through iterative reasoning, exploration, and tool-use, a process that can result in long, expensive context histories. While state-of-the-art \ac{se} agents like OpenHands or Cursor use LLM-based summarization to tackle this issue, it is unclear whether the increased complexity offers tangible performance benefits compared to simply omitting older observations.
  % Objective & Methods
  We present a systematic comparison of these approaches with SWE-agent on SWE-bench Verified across five diverse model configurations. Moreover, we show initial evidence of our findings generalizing to the OpenHands agent scaffold.
  % Contributions/Results
  We find that a simple environment Observation Masking strategy halves cost relative to the Raw Agent while matching, and sometimes slightly exceeding, the solve rate of LLM-Summary. Additionally, we introduce a novel hybrid approach that further reduces costs by 7\% and 11\% compared to just Observation Masking or LLM-Summary, respectively.
  % Conclusion
 Our findings raise concerns regarding the trend towards pure LLM-Summary and provide initial evidence of untapped cost reductions by pushing the efficiency-effectiveness frontier. 
 We release code and data for reproducibility.\footnote{Data: \url{https://huggingface.co/datasets/JetBrains-Research/the-complexity-trap}}\footnote{Code: \url{https://github.com/JetBrains-Research/the-complexity-trap}}
\end{abstract}

\section{Introduction}
The ambition to create autonomous agents that can independently handle complex \ac{se} tasks is rapidly becoming a reality. These agents, powered by \acp{llm}, typically operate in an iterative loop~\cite{yao_react_2023,wang_executable_2024}, at each turn they reason about the current state, devise a plan, and execute a tool (e.g., read a file, run tests). The output, or observation, from this tool is then added to the agent's context for the next turn, extending its problem-solving trajectory (\Cref{fig:trajectory-summarization-strategies-overview}). This agentic loop acts as a powerful test-time scaling mechanism~\cite{snell_scaling_test_time_2024,liu_test_time_1b_405b_2025,wu_inference_scaling_laws_2025}, utilizing the reasoning capabilities~\cite{wei_chain--thought_2023} of \acp{llm} at each turn while grounding them through environment responses.

\begin{wrapfigure}{l}{0.4\textwidth}
    \centering
    \includegraphics[width=\linewidth]{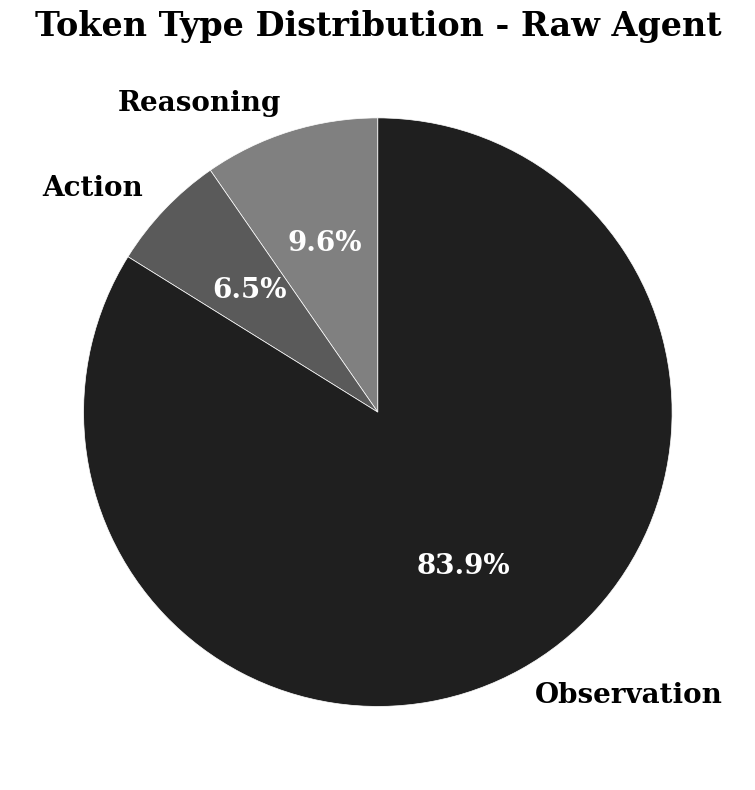}
    \caption{Environment observation tokens dominate the context window of an \ac{se} agent's trajectory.}
    \label{fig:token-type-distr}
\end{wrapfigure}

However, this iterative context expansion presents a fundamental tradeoff between cost and capability, or effectiveness and efficiency. As the agent's trajectory grows, calls to the \ac{llm} become prohibitively expensive due to token-based pricing, and inefficient due to the quadratic attention complexity in the wide-spread Transformer architecture~\cite{vaswani_attention_2017}. More critically, even with context windows exceeding 1M tokens, \acp{llm} suffer from the "lost in the middle" problem~\cite{liu_lost_in_the_middle_2024}. While \acp{llm} can process large context windows, they cannot properly make use of relevant information buried within their vast context~\cite{modarressi_nolima_2025,hong2025context}. This challenge is acutely amplified in the \ac{se} domain, where tool observations are notoriously verbose and noisy~\cite{lindenbauer-2025-ctim}. A single action can yield thousands of tokens, whether from reading an entire source file, running a recursive directory listing, or a lengthy test suite log. Concretely, observation tokens make up around $84\%$ of an average SWE-agent turn~\cite{yang_swe-agent_2024} (\Cref{fig:token-type-distr}) in our preliminary experiments (\Cref{appendix:preliminary-experiments}) on SWE-bench Lite-50~\cite{jimenez2024swebench,badertdinov2024scaling}. Due to this, targeting environment observations explicitly provides a strong baseline for \ac{llm}-agent context management.

Without any context management strategy targeting cost-efficiency, we find that agent costs can more than double (\Cref{tab:main_results}), making context management not just an optimization but an economic necessity. In fact, our results show that \textbf{any of the discussed management strategies are preferable to none}, as they consistently reduce costs and often improve performance. This raises a critical question: which strategy offers the best trade-off?

A natural baseline approach directly targets verbose environment observations through observation masking strategies that explicitly omit tool outputs. State-of-the-art systems like SWE-agent~\cite{yang_swe-agent_2024} and SWE-Search~\cite{antoniades2025swesearch} have adopted such relatively simple approaches. In parallel, an increasingly popular and more sophisticated solution is to prompt an \ac{llm} to perform trajectory summarization, replacing parts of the agent's history with condensed summaries (\Cref{fig:trajectory-summarization-strategies-overview}). This approach is adopted by prominent open-source and proprietary \ac{se} agents like OpenHands~\cite{wang2025openhands} or Cursor~\cite{cursor_summarization}. Open-source implementations of both the observation masking and \ac{llm} summarization approaches rely on heuristic triggers such as fixed context window sizes or turn thresholds. Thus, the key difference is whether they discard or condense old context. Despite the critical impact of this choice on agent cost and performance, the relative trade-offs between these approaches remain largely unexplored. 

In this work, we present a systematic comparison focused on the efficiency of context management strategies. For this, we analyze the performance of representative open-source Observation Masking and \ac{llm}-Summary implementations with respect to efficiency (cost in USD) and effectiveness (solve rate on the challenging and industry-standard SWE-bench Verified~\cite{openai_swe_bench_verified_2024} benchmark). We capitalize these names throughout to denote the specific strategies formally defined in \Cref{subsec:trajectory-management-strategies}. To enable controlled experimentation across model configurations, we implement \ac{llm}-Summary in SWE-agent~\cite{yang_swe-agent_2024} and adapt the OpenHands' \ac{llm}-Summary prompt (\Cref{fig:prompt:llm-summary}). We evaluate these strategies within the SWE-agent~\cite{yang_swe-agent_2024} scaffold and probe for generality on OpenHands~\cite{wang2025openhands}. Our experiments span model families, model sizes, licenses (open-weights vs. proprietary), and reasoning regimes (thinking vs. non-thinking). 

We find that both Observation Masking and \ac{llm}-Summary more than halve the cost compared to the Raw Agent. Furthermore, \ac{llm}-Summary is unable to consistently or significantly outperform the simple Observation Masking strategy across all model configurations. We show initial evidence of these findings generalizing to OpenHands. Furthermore, we introduce a novel hybrid approach that further reduces costs by 7\% and 11\% compared to just Observation Masking or \ac{llm}-Summary, respectively. These findings challenge current trends toward pure \ac{llm}-Summary and demonstrate that pure \ac{llm}-Summary strategies likely leave considerable cost savings untapped. 

\begin{figure}[t]
  \centering
  \includegraphics[width=\linewidth]{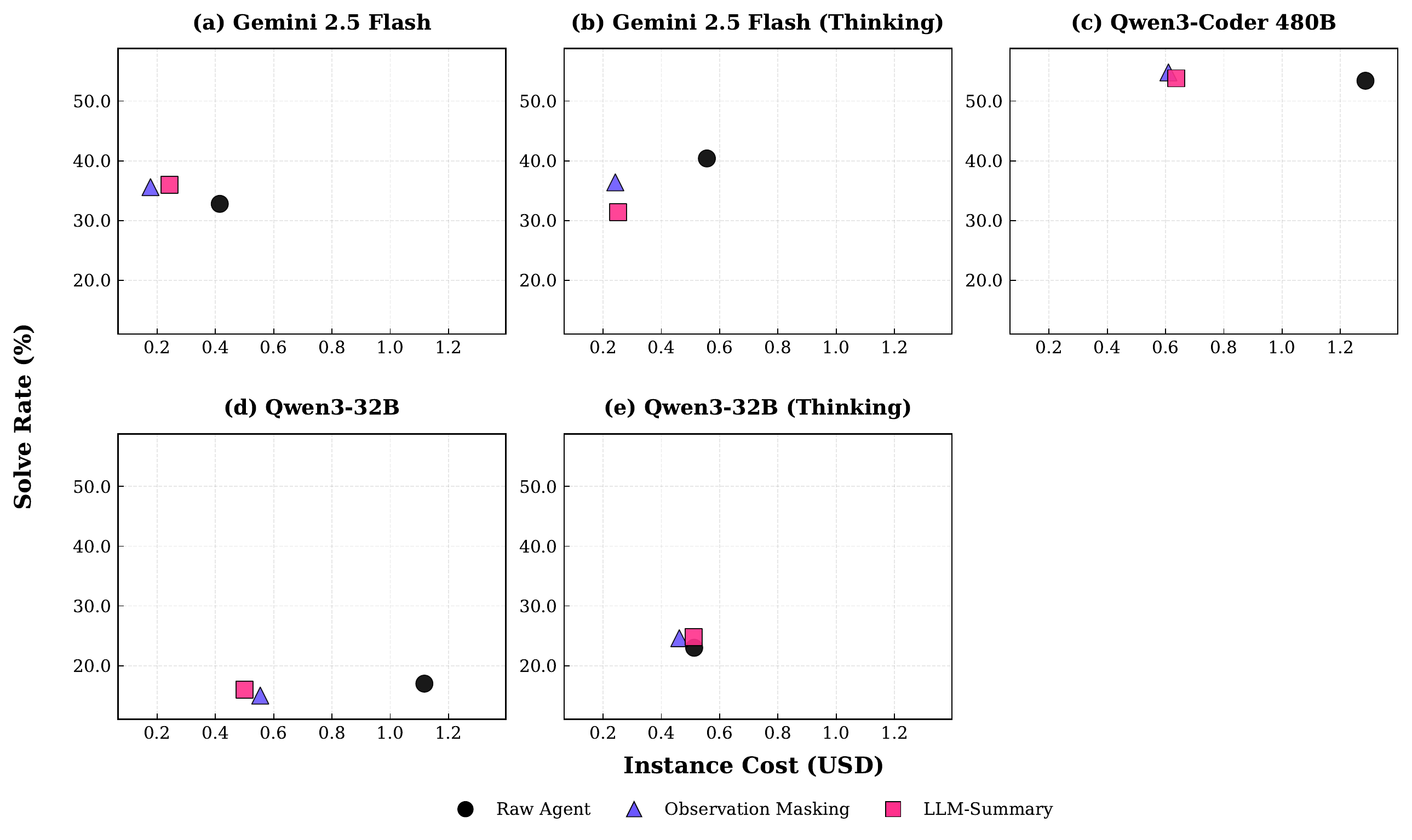}
  \caption{
    \textbf{The effectiveness versus efficiency tradeoff for context management strategies within SWE-agent~\cite{yang_swe-agent_2024} on SWE-bench Verified~\cite{openai_swe_bench_verified_2024}.} 
    The plot compares solve rate (y-axis, $\uparrow$) against the average cost per trajectory (x-axis, $\downarrow$) for different model configurations.
    We test each configuration with three strategies: Raw Agent (baseline, \textcolor{black}{\ding{108}}), \ac{llm}-Summary (\textcolor{jb_magenta}{\ding{110}}), and Observation Masking (\textcolor{jb_violet}{\ding{115}}). 
    Across all models, the Observation Masking strategy consistently occupies the most efficient frontier, achieving solve rates competitive with, and sometimes superior to, the \ac{llm}-Summary strategy. With Qwen3-Coder 480B~\cite{qwen3technicalreport,hui2024qwen2}, the best-performing model in our experiments, Observation Masking is not only 52\% cheaper than the Raw Agent baseline but also improves on the solve rate by 2.6 \%. Moreover, it even \textbf{reduces the cost per instance compared to \ac{llm}-Summary by \$0.03 (\$15 across 500 instances)}.
  }
  \label{fig:summarization-efficiency-efficacy}
\end{figure}

\section{Related Work}
\label{sec:related-work}
Current \ac{se} agent research mostly focuses on improving the effectiveness of \ac{se} agents by scaling training data~\cite{jain_r2e-gym,pan_swegym_2024,yang_swe-smith_2025}, selecting the most promising of multiple attempts~\cite{pan_swegym_2024,jain_r2e-gym,zhang2025diversity,antoniades2025swesearch,aggarwal-etal-2025-dars,xia_agentless_2024}, providing execution-free or execution-based feedback to the agent through critics~\cite{antoniades2025swesearch,pan_swegym_2024,jain_r2e-gym,zhang2025diversity,shinn_reflexion_2023}, or enhancing the agent's planning capabilities through explicit search strategies~\cite{antoniades2025swesearch,aggarwal-etal-2025-dars}. While these methods improve solve rate, they come at the cost of increased inference costs and thus reduced efficiency. Efficient context management for \ac{se} agents, on the other hand, has thus far received little attention. In this work, we investigate whether complex summarization strategies are necessary for efficient context management in \ac{se} agents. For this we experiment with SWE-agent~\cite{yang_swe-agent_2024} and OpenHands~\cite{wang2025openhands}.

Recently, several works have conducted in-depth analyses on the effect of the context size on the \acp{llm} performance~\cite{liu_lost_in_the_middle_2024,hong2025context,modarressi_nolima_2025}. These consistently show that \acp{llm} increasingly struggle to effectively utilize the context provided with increased context size. Despite the critical importance of context management for agent performance and deployment costs, existing work treats it as an implementation detail rather than a first-class research question. A recent exception, MEM1~\cite{zhou_mem1_2025}, explores dynamic state management for multi-hop QA~\cite{yang_hotpotqa_2018,naturalquestions2019kwiatkowski,jin2025search} and web navigation~\cite{yao_webshop_2023} tasks. However, the authors do not compare to omission-based approaches. Furthermore, the benchmarks used result in relatively short trajectories (hundreds of tokens)~\cite{zhou_mem1_2025} compared to \ac{se} agent trajectories that routinely are orders of magnitude larger~\cite{lindenbauer-2025-ctim}.

% collapse this to make space for CUA reference
Concurrently with our work, \citet{xiao_improving_2025} propose an \ac{llm}-Summary variant for efficient \ac{se} agent context management. However, they do not compare with an Observation Masking baseline. Most closely related, their ``Delete'' baseline is an \ac{llm}-Summary variant that deletes full turns instead of summarizing them. This baseline does not take advantage of \ac{se} trajectories skewing towards environment observations (\Cref{fig:token-type-distr}) and breaks the agent reasoning trace across turns, degrading agent downstream performance. Irrespective of this, their results show that this baseline is more efficient than \ac{llm}-Summary at a comparable downstream performance. Additionally, \citet{lu_scaling_2025} investigate using \ac{llm}-Summary for tackling context window limitations in RL training of \ac{se} and \ac{cua}. \citet{tang_beyond_2025} on the other hand, achieve impressive performance with Observation Masking for RL training and inference of deep research and \ac{cua} agents. This underscores the timeliness of our research and that our findings likely generalize to deep research and \ac{cua}.

\section{Experimental Configuration}
\label{sec:experimental-setup}
\begin{figure}[t]
  \centering
    \includegraphics[width=0.8\linewidth]{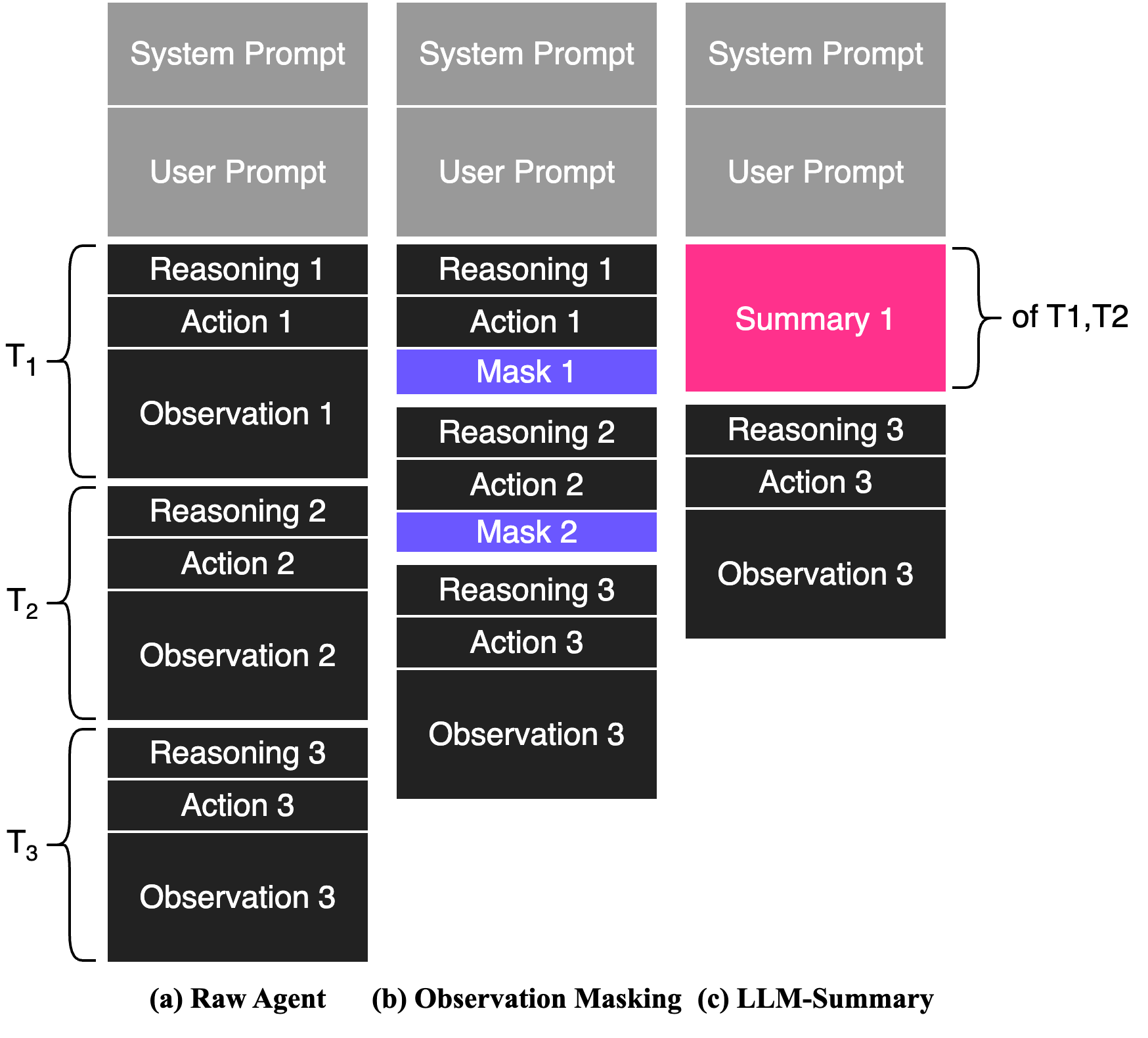}
  \caption{
    \textbf{Overview of the context management strategies evaluated in our work.} Box heights indicate the number of tokens in that portion of a typical trajectory.
    \textbf{(a)} The baseline ReAct-style~\cite{yao_react_2023} trajectory, where the context grows with each action-observation pair. 
    \textbf{(b)} \textbf{LLM-based summarization} condenses older turns into a running summary and preserving a few recent turns in full (e.g., OpenHands~\cite{wang2025openhands}).
    \textbf{(c)} \textbf{Observation masking} replaces observations older than a fixed window of M turns (here, M=1) with a placeholder (e.g., SWE-agent~\cite{yang_swe-agent_2024}).
  }
  \label{fig:trajectory-summarization-strategies-overview}
\end{figure}

In our experiments we investigate popular approaches to observation omission and \ac{llm}-based summarization through representative open-source implementations: (1) environment Observation Masking through a rolling window~\cite{yang_swe-agent_2024,antoniades2025swesearch}, and (2) prompt-based \ac{llm}-Summary~\cite{wang2025openhands,cursor_summarization}. In the following, we will define our chosen approaches in detail.

\subsection{Context Management Strategies}
\label{subsec:trajectory-management-strategies}
\paragraph{Raw Agent.}
The agent scaffolds we investigate use either ReAct~\cite{yao_react_2023} or CodeAct~\cite{wang_executable_2024} as an agent framework. In these frameworks, the agent's history, or trajectory, is a sequence of interactions with an environment. We formalize the trajectory at the end of turn $t-1$, denoted $\tau_{t-1}$, as a sequence of tokens:
\begin{equation}
\tau_{t-1} = (o_{sys}, o_{user}, (r_1, a_1, o_1), \dots, (r_{t-1}, a_{t-1}, o_{t-1}))
\end{equation}
where $o_{sys}$ and $o_{user}$ are the immutable system and user prompts that initialize the task (\Cref{fig:trajectory-summarization-strategies-overview}a). We define a turn $T_i = (r_i, a_i, o_i)$ consisting of reasoning $r_i$, action $a_i$ and observation $o_i$ as the atomic unit of agent-environment interaction at turn $i$. This allows us to compactly express $\tau_{t-1}$ as:
\begin{equation}
\tau_{t-1} = (o_{sys}, o_{user}, T_1, \dots, T_{t-1})
\end{equation}

At the beginning of turn $t$, the agent policy $\pi$, typically an \ac{llm}, conditions on the history $\tau_{t-1}$ to generate the next reasoning and action pair: $(r_t, a_t) \sim \pi(\cdot|\tau_{t-1})$. Without any context management strategy, $\tau$ grows with each turn, leading to excessive computational costs and eventual context length limitations.

\paragraph{Observation Masking.}
This strategy manages the context size by selectively condensing past environment observations while preserving the full history of agent reasoning and actions. We define an observation masking function, $f_{mask}(\tau_{t-1}, M)$, that takes the full trajectory $\tau_{t-1}$ and an integer window size $M$ as input. The function produces a condensed trajectory, $\tau'_{t-1}$, by replacing environment observations older than the window with a placeholder (\Cref{fig:trajectory-summarization-strategies-overview}b).

Formally, given the trajectory $\tau_{t-1} = (o_{sys}, o_{user}, T_1, \dots, T_{t-1})$, the transformed trajectory is:
\begin{equation}
\tau'_{t-1} = (o_{sys}, o_{user}, (r_1, a_1, o'_1), \dots, (r_{t-1}, a_{t-1}, o'_{t-1}))
\end{equation}
where the observation at each turn $i$, denoted $o'_i$, is conditionally defined as:
\begin{equation}
\label{eq:rolling_window}
o'_i = 
\begin{cases} 
      p_i & \text{if } i < t-M \\
      o_i & \text{if } i \geq t-M
\end{cases}
\end{equation}
Here, $p_{i}$ is a placeholder text representing the masked observation, such as ``Previous 8 lines omitted for brevity.''. In the following turns, the agent \ac{llm} then conditions on this condensed history $\tau'_{t-1}$ to produce $(r_t, a_t)$. This approach, following SWE-agent~\cite{yang_swe-agent_2024}, retains the complete reasoning chain while reducing distant observation fidelity. While this strategy reduces the speed at which the tokens in $\tau$ grow, it does not solve the issue of indefinite growth.

\paragraph{\ac{llm}-Summary.}
This strategy uses a "summarizer \ac{llm}" to condense the trajectory, which we denote as $\pi'$. In contrast to $f_{mask}$, the goal of this strategy is to maintain salient information of the processed turns. It is controlled by two parameters $N$ and $M$. $N$ regulates how many turns the agent will accumulate at once, and $M$ regulates how many trailing turns should be left unaltered. We trigger summarization when the accumulated turns since the last summary reach $N+M$. 

To help us define this approach, we will introduce two variables, $t_{last}$ and $s_{last}$. Let $t_{last}$ be the index of the final turn included in the most recent summary ($t_{last}=0$ at step 0). Let $s_{last}$ be the summary performed at index $t_{last}$ ($s_{last}=o_{user}$ at step 0). Then, we define the summarization as follows. First, we slice the trajectory to obtain the turns eligible for the summarization, containing the last summary $s_{last}$ and all turns between this summary and the $M$ to the last turn $\mathcal{T}_{sum} = (s_{last}, T_{t_{last}+1}, \dots, T_{t-1-M})$. Then, we generate a new summary $s_t$ by prompting $\pi'$ with a summary instruction $o_{si}$ and the relevant trajectory slice $\mathcal{T}_{sum}$.
\begin{equation}
\begin{split}
s_{t} &\sim \pi'(\cdot|o_{si}, \mathcal{T}_{sum})\\
t_{last} &= t-1-M
\end{split}
% \quad \text{where} \quad
% \mathcal{T}_{sum} = (s_{last}, T_{t_{last}+1}, \dots, T_{t-1-M})
\end{equation}
Finally, the history provided to the main policy $\pi$ is reconstructed into a new condensed trajectory, $\tau'_{t-1}$:
\begin{equation}
\label{eq:llm_summary_traj}
\tau'_{t-1} = (o_{sys}, o_{user}, s_t, T_{t-M}, \dots, T_{t-1})\end{equation}

On the next turn, the agent \ac{llm} conditions on this new, compact history: $(r_t, a_t) \sim \pi(\cdot|\tau'_{t-1})$. This method ensures the trajectory's growth is not only slowed, but bounded, as older interactions are recursively folded into an evolving summary.

\subsection{Experimental Configuration}
\label{subsec:experimental-configuration}
% This is models, infra etc
We conduct a rigorous, comparative study focusing on SWE-agent~\cite{yang_swe-agent_2024} and probing for generality with OpenHands~\cite{wang2025openhands} (\texttt{v0.43.0}). Our experiments cover diverse configurations spanning (1) model families, (2) model sizes, (3) model licenses (open-weights vs. proprietary), (4) reasoning regimes (thinking and non-thinking). Concretely, we use Qwen3-32B~\cite{qwen3technicalreport}\footnote{\url{https://huggingface.co/Qwen/Qwen3-32B}} in thinking and non-thinking mode with a context window of 122K tokens using YaRN~\cite{peng_yarn_2023}, Qwen3-Coder-480B-A35B-Instruct-FP8~\cite{qwen3technicalreport,hui2024qwen2}\footnote{\url{https://huggingface.co/Qwen/Qwen3-Coder-480B-A35B-Instruct-FP8}} with its default context window of 256K tokens, and Gemini 2.5 Flash~\cite{comanici_gemini_2025}\footnote{API Version: gemini-2.5-flash} with its default context window of 1M tokens in thinking and non-thinking mode. We conduct all experiments on SWE-bench Verified~\cite{openai_swe_bench_verified_2024} unless otherwise specified. We run all our experiments on a shared cluster of eight NVIDIA H200 GPUs, each equipped with 141 GB of HBM, and a total of 8 TB local disk storage. For the Qwen-32B~\cite{qwen3technicalreport} models we use two H200 GPUs, and 15 SWE-agent~\cite{yang_swe-agent_2024} inference workers. We choose a conservative number of workers, because we may encounter long trajectories with context sizes $>100$K tokens and must account for this in our KV-cache estimates. For Qwen3-Coder 480B~\cite{qwen3technicalreport,hui2024qwen2}, we use all eight GPUs and 35 inference workers. We use vLLM~\cite{kwon2023efficient} to serve the Qwen models on our cluster. In our experiments with Gemini we use eight inference workers on SWE-agent~\cite{yang_swe-agent_2024} and five on OpenHands~\cite{wang2025openhands} due to quota restrictions. For further details see \Cref{appendix:experimental-config}.

Our preliminary experiments (\Cref{appendix:preliminary-experiments}) indicate that the number of turns in the trajectory may influence the behavior of the context management strategies we investigate. Due to this, we experiment with long trajectories and thus set the turn limit to 250 in our experiments unless otherwise specified. For the Observation Masking strategy we use a rolling window size of $M=10$ in our main experiments, because it resulted in the best performance with SWE-agent~\cite{yang_swe-agent_2024} in our experiments (\Cref{appendix:ablation:window-size}). The \ac{llm}-Summary strategy we implement in SWE-agent~\cite{yang_swe-agent_2024} uses a slightly modified version of the OpenHands-style prompt (\Cref{appendix:llm-summary-prompt}). In contrast to OpenHands' baseline configuration~\cite{wang2025openhands}, we summarize 21 turns at once ($N=21$) and retain only the last ten ($M=10$) turns. Besides aligning the number of unaltered tail turns for the Observation Masking and \ac{llm}-Summary strategies, we also found that $M=10$ offered the best performance for the \ac{llm}-Summary strategy in our experiments (\Cref{appendix:ablation:tail-length}). For the agent model we use a temperature of 0.8, and for the summary model a temperature of zero. In contrast to experiments with Qwen3-32B~\cite{qwen3technicalreport} thinking, we restrict the thinking budget of Gemini 2.5 Flash to zero or 800 tokens (denoted as \textit{thinking}) due to cost constraints. 

\section{Main Results}
\label{sec:main-results}
Our main experiments within SWE-agent~\cite{yang_swe-agent_2024} evaluate three context management strategies, with results summarized in Table \ref{tab:main_results}. The results reveal two central findings that hold robustly across the diverse conditions we tested. First, both Observation Masking and \ac{llm}-Summary significantly reduce costs, without significantly reducing solve rate performance. Second, \ac{llm}-Summary does not consistently, or significantly outperform Observation Masking on efficiency or effectiveness. For further details see \Cref{appendix:main-resutlts-details}.

\subsection{The Universal Benefit of Context Management}
\label{subsec:universal-benefit}
Our first and most foundational finding, reinforcing the motivation of this study, is that context management is not merely an optimization but a necessity. As shown in \Cref{tab:main_results}, leaving the agent's context to grow unchecked (the Raw Agent baseline) is consistently the most expensive strategy. In all experimental configurations where trajectories are long enough to benefit from efficient context management, both Observation Masking and \ac{llm}-Summary significantly reduce the cost per instance, in most cases by more than 50\%. We discuss the outlying behavior of Qwen3-32B (thinking) in \Cref{sec:short-traj-qwen32-thinking}.

Furthermore, this efficiency does not necessarily come at the cost of performance. In three of our five setups, the most efficient strategy also achieved a higher solve rate than the Raw Agent baseline. This demonstrates that beyond a certain point, more context becomes a liability rather than an asset, aligning with the "lost in the middle" problem~\cite{liu_lost_in_the_middle_2024}. Therefore it is critical to question which approach offers the best trade-off between effectiveness and efficiency.

\newcommand{\good}[1]{\textcolor{ForestGreen}{#1}}
\newcommand{\bad}[1]{\textcolor{red}{#1}}
\newcommand{\sig}{\textsuperscript{†}}

\begin{table}[ht]
\caption{\textbf{Comparison of context management strategies with 95\% bootstrap confidence intervals.} We report change and significance (†) compared to the \textit{Raw Agent}. We report Solve Rate (effectiveness, $\uparrow$) and Instance Cost (efficiency, $\downarrow$). For each model, we \textbf{boldface} the best-performing context management strategy for each metric. All experiments use SWE-agent~\cite{yang_swe-agent_2024} on SWE-bench Verified~\cite{openai_swe_bench_verified_2024}. Further details in \Cref{appendix:main-resutlts-details}.}
\label{tab:main_results}
\centering
\resizebox{\textwidth}{!}{
\begin{tabular}{llll}
\toprule
\textbf{Model} & \textbf{Strategy} & \textbf{Solve Rate (\%,$\uparrow$)} & \textbf{Instance Cost (\$,$\downarrow$)} \\
\midrule
\multirow{3}{*}{Qwen3-32B}
                & Raw Agent           & 17.0\,$\pm3.3$ & 1.12\,$\pm0.18$ \\
                & Observation Masking & 15.0\,$\pm3.1$ (\bad{-11.8\%}) & 0.55\,$\pm0.09$ (\good{-50.9\%})\sig \\
                & LLM-Summary         & \textbf{16.0}\,$\pm3.3$ (\bad{-5.9\%}) & \textbf{0.50}\,$\pm0.07$ (\good{-55.4\%})\sig \\
\midrule
\multirow{3}{*}{\parbox{3.5cm}{Qwen3-32B \\ (thinking)}} 
                & Raw Agent           & 23.0\,$\pm3.7$ & 0.51\,$\pm0.07$ \\
                & Observation Masking & 24.6\,$\pm3.8$ (\good{+7.0\%}) & \textbf{0.46}\,$\pm0.05$ (\good{-9.8\%}) \\
                & LLM-Summary         & \textbf{24.8}\,$\pm3.9$ (\good{+7.3\%}) & 0.51\,$\pm0.06$ (0.0\%) \\
\midrule
\multirow{3}{*}{Qwen3-Coder 480B} 
                & Raw Agent           & 53.4\,$\pm4.3$ & 1.29\,$\pm0.26$ \\
                & Observation Masking & \textbf{54.8}\,$\pm4.4$ (\good{+2.6\%}) & \textbf{0.61}\,$\pm0.06$ (\good{-52.7\%})\sig \\
                & LLM-Summary         & 53.8\,$\pm4.2$ (\good{+0.7\%}) & 0.64\,$\pm0.06$ (\good{-50.4\%})\sig \\
\midrule
\multirow{3}{*}{Gemini 2.5 Flash} 
                & Raw Agent           & 32.8\,$\pm4.1$ & 0.41\,$\pm0.08$ \\
                & Observation Masking & 35.6\,$\pm4.2$ (\good{+8.5\%}) & \textbf{0.18}\,$\pm0.03$ (\good{-56.1\%})\sig \\
                & LLM-Summary         & \textbf{36.0}\,$\pm4.1$ (\good{+9.8\%}) & 0.24\,$\pm0.04$ (\good{-41.5\%})\sig \\
\midrule
\multirow{3}{*}{\parbox{3.5cm}{Gemini 2.5 Flash \\ (thinking)}} 
                & Raw Agent           & 40.4\,$\pm4.3$ & 0.56\,$\pm0.10$ \\
                & Observation Masking & \textbf{36.4}\,$\pm4.2$ (\bad{-9.9\%})\sig & \textbf{0.24}\,$\pm0.04$ (\good{-57.1\%})\sig \\
                & LLM-Summary         & 31.4\,$\pm4.0$ (\bad{-22.3\%})\sig & 0.25\,$\pm0.05$ (\good{-55.4\%})\sig \\
\bottomrule
\end{tabular}
}
\end{table}

\subsection{Observation Masking: Dominant Efficiency with Minimal Complexity}
Having established the clear need for context management, we now turn to the second central finding of our work: the surprising power of simplicity. As we can see in \Cref{tab:main_results}, in four out of five experimental setups Observation Masking yielded the lowest cost per instance. It achieves this by drastically reducing the number of environment observation tokens processed in each agent turn without incurring the computational overhead of a separate summarization call. This is highly effective, because the agent's context skews heavily towards environment observations in \ac{se} (\Cref{fig:token-type-distr}). Furthermore, this strategy requires fewer warm-up turns than \ac{llm}-Summary ($M=10$ vs $N+M=31$). This results in quicker and more robust cost reductions, even on short trajectories (e.g., Qwen3-32B (thinking), see \Cref{sec:short-traj-qwen32-thinking}).

Notably, this finding holds across model configurations. Furthermore, while \$0.03 in cost reductions between \ac{llm}-Summary and Observation Masking for Qwen3-Coder 480B seems small, it already amounts to \$15 across the entire benchmark. This highlights that even small cost-efficiency gains can have a significant impact on the economic viability or large-scale \ac{llm} agent deployments and underscores the need for research on efficient context management.

\subsection{Challenging the Need for Complex Summaries}
Beyond consistently being the cheapest option, Observation Masking proves to be remarkably effective at maintaining high solve rates. This directly challenges the assumption that complex, semantic summarization is necessary to preserve critical information from an agent's trajectory.

In fact, Observation Masking not only competes with \ac{llm}-Summary, but can outperform it. With the Qwen3-Coder 480B~\cite{qwen3technicalreport,hui2024qwen2} model, Observation Masking achieved a solve rate of 54.8\%, a slight improvement over the \ac{llm}-Summary's 53.8\%. Similarly, for Gemini 2.5 Flash~\cite{comanici_gemini_2025} (thinking), it outperformed \ac{llm}-Summary by five percentage points. In the cases where LLM-Summary did perform better, such as with Gemini 2.5 Flash, the margin was minimal (36.0\%\ vs.\ 35.6\%). This indicates that Observation Masking consistently performs on-par with, or better than the \ac{llm}-Summary strategy.

The implication is clear: the most recent context is often sufficient for \ac{se} agents. Retaining the entire history, or even a sophisticated summary of it, may not be the most effective use of the model's limited context window and our research budget.

\subsection{The Trajectory Elongation Effect}
\label{subsec:main-results:elongation}
A key question arising from our main results in \Cref{tab:main_results} is why the \ac{llm}-Summary strategy is less cost-effective than the Observation Masking strategy for all experiments, except Qwen3-32B. Our analysis reveals that this partially stems from an unexpected "trajectory elongation" side-effect of the \ac{llm}-Summary context management strategy.

\begin{figure}[t]
    \centering
    \includegraphics[width=0.9\linewidth]{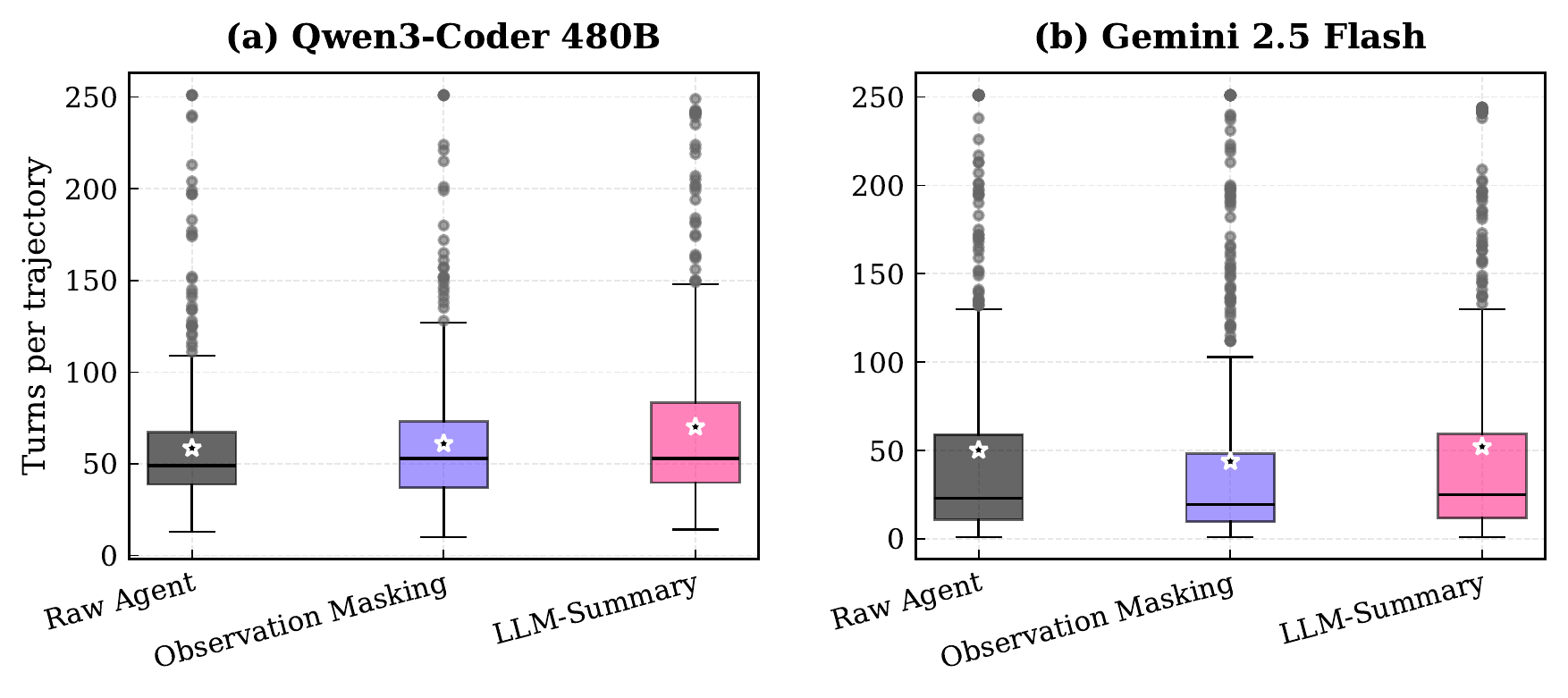}
    \caption{\textbf{Impact of context management strategies on trajectory length.} Box plots show the distribution of trajectory lengths (in turns) across different strategies within SWE-agent~\cite{yang_swe-agent_2024}. \ac{llm}-Summary consistently leads to longer trajectories, suggesting they mask failure signals that would otherwise prompt earlier termination. The star indicates the mean trajectory length.}
    \label{fig:trajectory_lengths_boxplots}
\end{figure}

For this, we analyze the distribution of turns per instance in \Cref{fig:trajectory_lengths_boxplots}. We note that \ac{llm}-Summary leads to longer mean trajectory lengths for both Qwen3-Coder 480B and Gemini 2.5 Flash. For Gemini 2.5 Flash the mean trajectory length using \ac{llm}-Summary is 52 turns, which is a $15\%$ increase over the mean trajectory length of the Observation Masking (44 turns) and $4\%$ over that of the Raw Agent (50 turns). Likewise, for Qwen3-Coder 480B, we observe an increase of the mean trajectory length by $15\%$ compared to the Raw Agent and $13\%$ compared to the Observation Masking strategy.

Notably, this trend of turn elongation translates directly to the efficiency of a strategy observed in \Cref{fig:summarization-efficiency-efficacy} and \Cref{tab:main_results}. The only experiment for which the \ac{llm}-Summary strategy proved more cost-efficient is Qwen3-32B. In this case, the Observation Masking strategy, rather than the \ac{llm}-Summary strategy led to a $13\%$ increase in mean trajectory length compared to the Raw Agent. This indicates that context summaries act as a reinforcing signal, encouraging the agent to keep going. This in turn results in trajectory elongation that diminishes the efficiency gained through the bounded context achieved by \ac{llm}-Summary (\Cref{subsec:trajectory-management-strategies}).

\section{Discussion}
\label{sec:discussion}
In this section we probe the generality of our findings on OpenHands~\cite{wang2025openhands}, discuss the impact of the \ac{llm} summary generation on the cost structure of the \ac{llm}-Summary strategy, and introduce a novel hybrid approach combining the strengths of both strategies.

\subsection{Probing for Generality With OpenHands}
\label{subsec:openhands}
To investigate the generality of our main results across scaffolds, we probe for generality with OpenHands on a 50-instance slice of SWE-bench Verified~\cite{openai_swe_bench_verified_2024,badertdinov2024scaling} using Gemini 2.5 Flash~\cite{comanici_gemini_2025} (no thinking), with a turn limit of 250, \ac{llm}-Summary N=21, M=10, and Observation Masking M=10 and M=58. We present the results \Cref{fig:combined-tradeoff}a. 

First, we note that the Observation Masking rolling window size $M$ is an agent-specific hyperparameter that requires tuning. If we simply re-use the optimal value from SWE-agent~\cite{yang_swe-agent_2024}, Observation Masking performance degrades drastically. However, after tuning, we can reproduce our results on this agent scaffold. We hypothesize that we need to tune this hyperparameter due to scaffold-specific implementation details. For example, SWE-agent~\cite{yang_swe-agent_2024} directly elides retries due to syntax errors from the dialog history. However, OpenHands~\cite{wang2025openhands} retains such retry turns. This means we need a larger window size to retain an informative context for this agent. 

Overall, this provides initial evidence that our findings generalize across agent scaffolds if the agent's context similarly skews toward environment observations, as typically is the case in \ac{se} agents.

\subsection{The Costs of Summarization}
\label{subsec:cost-summary}
A closer examination of the cost breakdown reveals that the efficiency gap between strategies stems from two complementary effects. 
First, the ``trajectory elongation'' effect discussed in \Cref{subsec:main-results:elongation}, and second, the costs of generating the summary.

As shown in \Cref{tab:summary_instance_costs}, the direct API cost of generating summaries accounts for up to 7.2\% of the total instance cost. Importantly, these summarization calls are particularly expensive because each requires processing a unique sequence of turns, limiting cache reuse to the \ac{llm}-Summary system prompt (\Cref{fig:prompt:llm-summary}). This poses a critical limitation given that, several modern \ac{llm} APIs (e.g., Gemini) offer substantially cheaper cache hits than cache misses (up to $10\times$ cheaper). Once we subtract these summarization costs from the total, the efficiency difference between \ac{llm}-Summary and Observation Masking largely disappears for most experiments. This indicates that the summarization API calls themselves constitute a substantial portion of the efficiency gap. Nonetheless, the more complex \ac{llm}-Summary strategy is still unable to significantly or consistently outperform the remarkably strong Observation Masking strategy on cost-efficiency. This hints at potentially untapped cost savings through underexplored Observation Masking or hybrid approaches.

\begin{table}[ht]
  \caption{\textbf{Mean Instance \ac{llm}-Summary Cost per Model.} \ac{llm} summary generation API costs explain a portion of the cost-efficiency difference between the \ac{llm}-Summary and Observation Masking strategy.}
  \label{tab:summary_instance_costs}
  \centering
  \begin{tabular}{l S S[table-format=1.6]}
    \toprule
    \textbf{Model} & {\textbf{Instance \ac{llm}-Summary Cost (\$)}} & {\textbf{Proportional Cost (\%)}}\\
    \midrule
    Qwen3-32B & 0.0143 & 2.86\\
    Qwen3-32B (thinking) & 0.0033 & 0.65\\
    Qwen3-Coder 480B & 0.0439 & 7.20\\
    Gemini 2.5 Flash & 0.0161 & 6.71\\
    Gemini 2.5 Flash (thinking) & 0.0131 & 5.24 \\
    \bottomrule
  \end{tabular}
\end{table}

\subsection{Hybrid: Combined Observation Masking and LLM-Summary}
\label{subsec:discussion:hybrid}
\begin{figure}[t]
    \centering
    \begin{subfigure}[t]{0.48\linewidth}
        \centering
        \includegraphics[width=\linewidth]{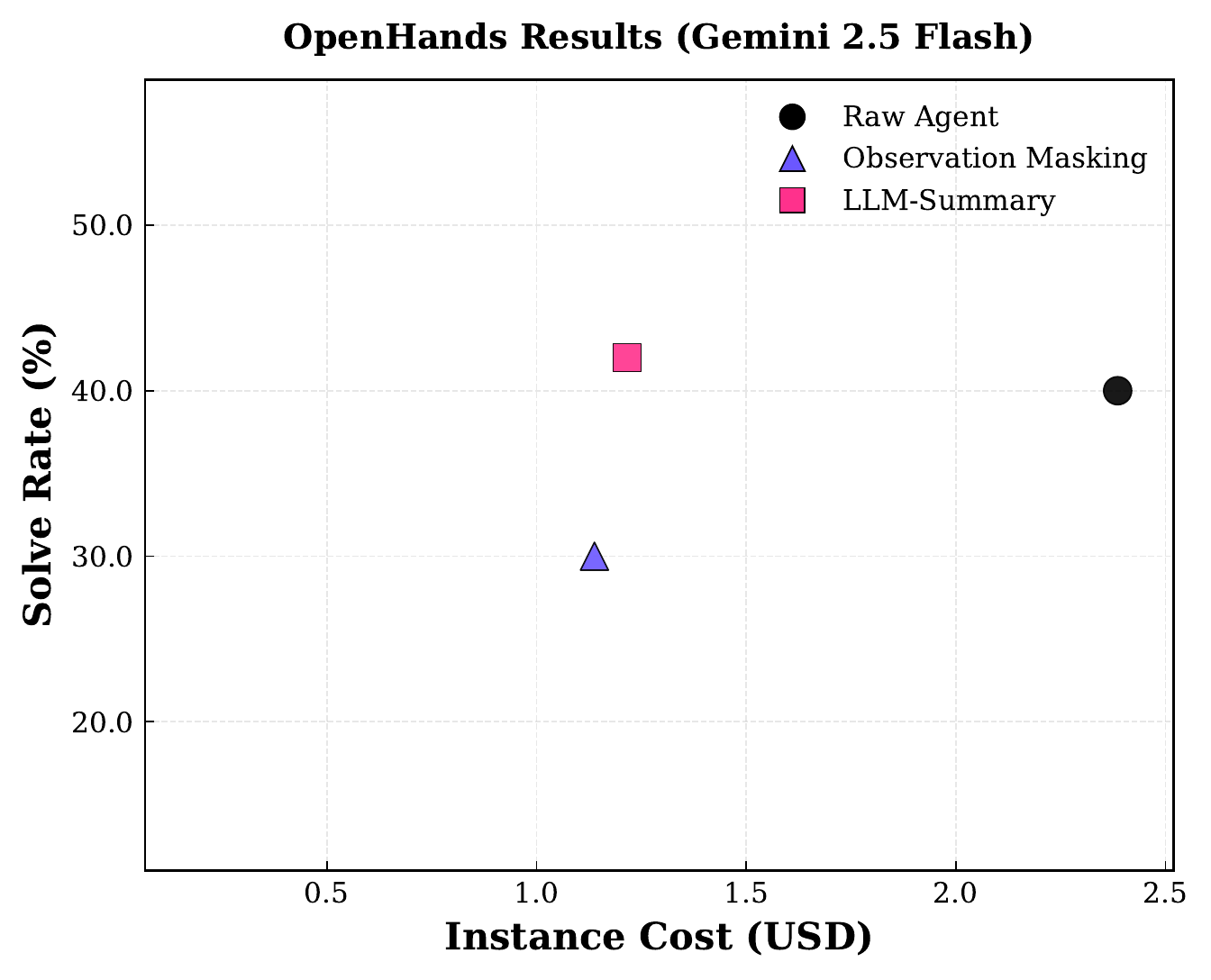}
        \label{fig:openhands-scatter}
    \end{subfigure}
    \hfill
    \begin{subfigure}[t]{0.48\linewidth}
        \centering
        \includegraphics[width=\linewidth]{data/qwen3_coder_single_scatter.png}
        \label{fig:exploratory-hybrid}
    \end{subfigure}
    \caption{\textbf{(a) Probing the generality of our findings with OpenHands~\cite{wang2025openhands} on the SWE-bench Verified-50~\cite{badertdinov2024scaling} subset}. After appropriately tuning the rolling window size $M$ to the agent scaffold, Observation Masking again matches the performance of \ac{llm}-Summary on both cost and solve rate. \textbf{(b) Our novel hybrid Observation Masking and LLM-Summary approach on SWE-bench Verified-50~\cite{badertdinov2024scaling} on the SWE-agent scaffold using Qwen3-Coder 480B.} Effectively combining the strengths of each approach results in a strategy that robustly realizes efficiency gains regardless of trajectory length while benefiting from bounded context on excessively long trajectories. Our hybrid approach yields a slight solve-rate gain of 2.6 percent points compared to the Raw Agent while reducing costs by 7\% and 11\% compared to Observation Masking and \ac{llm}-Summary, respectively.}
    \label{fig:combined-tradeoff}
\end{figure}

Motivated by the strong individual performances of the context management strategies we cover, both significantly and consistently reducing cost by $>50\%$, we present a novel hybrid approach in this section. For this, we experiment with the strongest model with respect to solve rate we cover, Qwen3-Coder 480B with SWE-agent~\cite{yang_swe-agent_2024} on SWE-bench Verified-50~\cite{badertdinov2024scaling} due to cost reasons. We visualize the results in \Cref{fig:combined-tradeoff}b.

Both the trajectory elongation effect (\Cref{subsec:main-results:elongation}) and the summary generation overhead (\Cref{subsec:cost-summary}) motivate us to treat \ac{llm}-Summary as a last resort strategy and defer it as long as possible. However, if we increase $N$, we increase the number of warm-up turns needed until we observe any effect. During this accumulation phase, we operate under the costly Raw Agent regime. Moreover, we risk not observing any effects on short to medium length trajectories (\Cref{sec:short-traj-qwen32-thinking}). Using Observation Masking during the turn accumulation phase allows us to combine the strengths of each approach. By increasing $N$, we defer \ac{llm}-Summary and treat this approach as a last resort for bounding the context of long trajectories. At the same time, Observation Masking quickly realizes gains during turn accumulation. Moreover, it does so robustly even on short trajectories.

We set $N=43$, because at this number of turns the context accumulated under the Observation Masking regime approximately matches the context accumulated under the Raw Agent at $N=21$ turns ($\approx30K$ tokens, see \Cref{fig:preliminary_experiments}). To avoid notation clash, we use $W$ for the rolling window size of Observation Masking in the hybrid setup. Overall, we use $N=43,M=W=10$ for the hybrid setup. Note that we pass the unmasked context whenever summarizing with \ac{llm}-Summary.

Compared to Observation Masking and \ac{llm}-Summary, this approach reduces costs by 7\% and 11\%, respectively. Moreover, it even improves the downstream task performance by 2.6 percent points, pushing the effectiveness-efficiency frontier. This results in expected savings of \$20 compared to Observation Masking and \$35 compared to \ac{llm}-Summary on the full SWE-bench Verified~\cite{openai_swe_bench_verified_2024} benchmark.

To ablate our hyperparameter choice, we also test the hybrid approach with a naive choice of hyperparameters, disregarding any strategy-specific properties. For this, we simply re-use the hyperparameters from the individual approaches $N=21,M=W=10$. With this configuration, the hybrid approach actually degrades the systems' overall cost efficiency, due to compounding KV cache inefficiencies, and cost overhead due to \ac{llm}-Summary invocations.

\section{Limitations}
\label{sec:limitations}
While our study provides a rigorous evaluation of context management strategies, its scope has three main limitations. First, we experiment exclusively within the \ac{se} domain, using the SWE-bench~\cite{jimenez2024swebench} benchmark. This domain is characterized by long, verbose tool outputs, a condition that naturally favors the efficiency of Observation Masking. Consequently, our findings on the superiority of this strategy may not generalize to domains where agent-environment interactions are more succinct.
Second, all strategies investigated use simple, non-adaptive heuristic triggers. Observation Masking employs a fixed-size rolling window that is agnostic to the relevance or staleness of past observations (e.g., retaining a file's content after it was modified). Similarly, \ac{llm}-Summary operates on a fixed turn-based schedule, ignoring semantic boundaries or agent subgoals. Finally, while we provide initial evidence of generalization across agent scaffolds, a more comprehensive investigation may be warranted.

\section{Conclusion}
\label{sec:conclusion}
This work presents a comprehensive study on context management strategies spanning diverse model configurations and agent scaffolds. We find that efficient context management strategies consistently and significantly reduce system costs by $>50\%$ without significantly reducing downstream performance. Surprisingly, the popular \ac{llm}-Summary strategy is unable to consistently or significantly outperform the simple Observation Masking baseline. This hints at untapped savings potential in modern agentic systems that focus only on \ac{llm}-Summary. We empirically validate this hypothesis with our novel hybrid Observation Masking and \ac{llm}-Summary strategy that further reduces costs by 7\% and 11\% compared to Observation Masking and \ac{llm}-Summary while improving the downstream task performance. These findings establish the critical need for context management to enable economically feasible, and environmentally sustainable \ac{llm} agent deployment. Moreover, it highlights that in the quest for efficient \ac{llm} agents, simple solutions can be surprisingly effective.

\begin{ack}
We would like to thank Calvin Smith for the productive and encouraging discussions on the context management strategies implemented in OpenHands. We would also like to thank Kirill Gelvan for his feedback on early versions of this paper.
\end{ack}

\bibliography{custom}

\appendix
\section{Experimental Configuration}
\label{appendix:experimental-config}
To compute the instance cost reported in our experiments, we distinguish between proprietary models we access through the API, and models we host locally on our infrastructure. For the Gemini~\cite{comanici_gemini_2025} experiments, which we access through the Vertex AI API, we report the cost returned by the API. For Qwen~\cite{qwen3technicalreport,hui2024qwen2} experiments, we self-host the models as described in \Cref{subsec:experimental-configuration}. To compute the cost, we use the observed per-turn token usage and post-hoc compute the cost based on the official Alibaba API pricing\footnote{ \url{https://www.alibabacloud.com/help/en/model-studio/models\#16ff9753e1ctz}}. Note that in contrast to the Vertex AI platform, the official Alibaba API pricing for the Qwen3-32B model, does not distinguish between cache hit and miss input tokens. This leads to an inflated cost of the Qwen3-32B experiments. We refer readers interested in the cost structure of our experiments under different pricing schemes to out HuggingFace dataset\footnote{Data: \url{https://huggingface.co/datasets/JetBrains-Research/the-complexity-trap}}, in which we release all experimental data for our main experiments and our repository\footnote{Code: \url{https://github.com/JetBrains-Research/the-complexity-trap}}.

\definecolor{promptlavender}{RGB}{230, 230, 250}
\definecolor{promptlavenderaccent}{RGB}{99,99,224}

\section{Short Trajectories in Qwen3-32B (thinking): Hypotheses and Implications}
\label{sec:short-traj-qwen32-thinking}
The Qwen3-32B (thinking) configuration exhibits an around 50\% shorter median trajectory length compared to the Qwen3-32B configuration. Given that the only change between the two configurations is that we enable model thinking, this is surprising. In the following, we detail our investigation of this outlier result and discuss its implications on the interpretation of our findings.

\paragraph{Configuration validation.} We verified that the configuration and dataset are correct for these two experimental configurations across all context management strategies. Additionally, we did not observe deviations in the distribution of instance exit statuses between the two configurations, besides the slightly improved performance of the model under the thinking regime (\Cref{chapter:results}). 

\paragraph{Qualitative analysis.} As we did not identify a misconfiguration, we further performed a qualitative analysis of the same 20 (4\%) trajectories across the Qwen3-32B experimental configurations. Here, we noticed that both models sometimes struggle with following the function calling format of the agent scaffold. However, as discussed above, we did not observe suspicious deviations in the number of exits due to function calling errors in either configuration.

\paragraph{Implications.} Because we could not identify any misconfiguration or otherwise suspicious behavior as the reason for this outlier result with the Qwen3-32B (thinking) configuration, we must assume that it is valid. Thus, we now discuss the implications of this result on the interpretation of our overall findings. First, recall that this configuration resulted in short trajectories. However, both of our context management strategies need a number of warm-up turns, before they start modifying the trajectory and thus reducing cost. For \ac{llm}-Summary we require $N+M=31$ turns before we produce the first summary with our hyperparameters. This means the trajectories with a median length of 15 turns are far too short to realize efficiency gains from this context management strategy. Observation Masking on the other hand, starts masking observations after $M=10$ turns. This means we expect to see an effect when using this strategy on shorter trajectories, however it may be muted. This exactly matches the empirical behavior observed in \Cref{chapter:results}. Therefore, we attribute the insignificant cost savings in the Qwen3-32B (thinking) configuration to the shorter trajectory lengths, rather than fundamental issues with our context management strategies. This interpretation is further supported by the fact that even in this unfavorable setting, Observation Masking reduces cost by $\approx10\%$, and both strategies still result in stable downstream performance.

\section{Detailed Main Results}
\label{appendix:main-resutlts-details}
In this section, we provide further data on which we base our confidence intervals and significance indicators in \Cref{tab:main_results}. Table~\ref{tab:main_results-detailed} presents the full asymmetric confidence intervals underlying our main results. The symmetric intervals in Table~\ref{tab:main_results} are derived by averaging the asymmetric bounds: e.g., $17.0_{-3.2}^{+3.4}$ yields $17.0 \pm 3.3$.

\begin{table}[ht]
\caption{Comparison of context management strategies with 95\% bootstrap confidence intervals, showing asymmetry. We use † to indicate significance compared to the Raw Agent. We report Solve Rate (effectiveness, $\uparrow$) and Instance Cost (efficiency, $\downarrow$). For each model, we \textbf{boldface} the best-performing context management strategy for each metric (relative to the Raw Agent baseline). Change is reported relative to the \textit{Raw Agent} baseline. All experiments use SWE-agent~\cite{yang_swe-agent_2024} on SWE-bench Verified~\cite{openai_swe_bench_verified_2024}.}
\label{tab:main_results-detailed}
\centering
\resizebox{\textwidth}{!}{
\begin{tabular}{llll}
\toprule
\textbf{Model} & \textbf{Strategy} & \textbf{Solve Rate (\%,$\uparrow$)} & \textbf{Instance Cost (\$,$\downarrow$)} \\
\midrule
\multirow{3}{*}{Qwen3-32B}
& Raw Agent           & 17.0\,${}_{-3.2}^{+3.4}$ & 1.12\,${}_{-0.17}^{+0.18}$ \\
& Observation Masking & 15.0\,${}_{-3.0}^{+3.2}$ (\bad{-11.8\%}) & 0.55\,${}_{-0.08}^{+0.09}$ (\good{-50.9\%})\sig \\
& LLM-Summary         & \textbf{16.0}\,${}_{-3.2}^{+3.4}$ (\bad{-5.9\%}) & \textbf{0.50}\,${}_{-0.06}^{+0.07}$ (\good{-55.4\%})\sig \\
\midrule
\multirow{3}{*}{\parbox{3.5cm}{Qwen3-32B \\ (thinking)}} 
                & Raw Agent           & 23.0\,${}_{-3.6}^{+3.8}$ & 0.51\,${}_{-0.06}^{+0.07}$ \\
                & Observation Masking & 24.6\,${}_{-3.8}^{+3.8}$ (\good{+7.0\%}) & \textbf{0.46}\,${}_{-0.05}^{+0.05}$ (\good{-9.8\%}) \\
                & LLM-Summary         & \textbf{24.8}\,${}_{-3.8}^{+4.0}$ (\good{+7.3\%}) & 0.51\,${}_{-0.06}^{+0.06}$ (0.0\%) \\
\midrule
\multirow{3}{*}{Qwen3-Coder 480B} 
                & Raw Agent           & 53.4\,${}_{-4.2}^{+4.4}$ & 1.29\,${}_{-0.24}^{+0.28}$ \\
                & Observation Masking & \textbf{54.8}\,${}_{-4.4}^{+4.4}$ (\good{+2.6\%}) & \textbf{0.61}\,${}_{-0.05}^{+0.06}$ (\good{-52.7\%})\sig \\
                & LLM-Summary         & 53.8\,${}_{-4.2}^{+4.2}$ (\good{+0.7\%}) & 0.64\,${}_{-0.05}^{+0.06}$ (\good{-50.4\%})\sig \\
\midrule
\multirow{3}{*}{Gemini 2.5 Flash} 
                & Raw Agent           & 32.8\,${}_{-4.0}^{+4.2}$ & 0.41\,${}_{-0.07}^{+0.08}$ \\
                & Observation Masking & 35.6\,${}_{-4.2}^{+4.2}$ (\good{+8.5\%}) & \textbf{0.18}\,${}_{-0.02}^{+0.03}$ (\good{-56.1\%})\sig \\
                & LLM-Summary         & \textbf{36.0}\,${}_{-4.0}^{+4.2}$ (\good{+9.8\%}) & 0.24\,${}_{-0.04}^{+0.04}$ (\good{-41.5\%})\sig \\
\midrule
\multirow{3}{*}{\parbox{3.5cm}{Gemini 2.5 Flash \\ (thinking)}} 
                & Raw Agent           & 40.4\,${}_{-4.4}^{+4.2}$ & 0.56\,${}_{-0.10}^{+0.10}$ \\
                & Observation Masking & \textbf{36.4}\,${}_{-4.2}^{+4.2}$ (\bad{-9.9\%}) \sig & \textbf{0.24}\,${}_{-0.03}^{+0.04}$ (\good{-57.1\%})\sig \\
                & LLM-Summary         & 31.4\,${}_{-4.0}^{+4.0}$ (\bad{-22.3\%})\sig & 0.25\,${}_{-0.04}^{+0.05}$ (\good{-55.4\%})\sig \\
\bottomrule
\end{tabular}
}
\end{table}

\subsection{Statistical Analysis}
We assess significance using paired nonparametric bootstrap with $B = 10{,}000$ replicates and show detailed results in \Cref{tab:appendix_pvals_bootstrap}. For each model-strategy pair, we compute the paired difference $\Delta = \text{mean(strategy)} - \text{mean(raw)}$ on the same $n = 500$ instances, preserving instance-level correlations. We report:
\begin{itemize}
    \item 95\% percentile confidence intervals
    \item Two-sided p-values: $p = 2 \times \min(\Pr(\Delta^* \geq 0), \Pr(\Delta^* \leq 0))$
    \item Significance markers (†) when $p < 0.05$
\end{itemize}

Table~\ref{tab:appendix_pvals_bootstrap} provides the complete bootstrap statistics. Note that p-values of 0.0000 indicate no sign-crossing across all bootstrap replicates (resolution $\leq 10^{-4}$).

\begin{table}[ht]
\caption{Paired bootstrap differences vs. Raw Agent with 95\% percentile CIs and two-sided bootstrap p-values (B=10{,}000). $\Delta$ Solve Rate is reported in percentage points (pp), $\Delta$ Mean Cost in dollars per instance. Negative cost differences indicate cost savings. All rows use n=500 common instances per model. We use † to indicate significance compared to the Raw Agent.}
\label{tab:appendix_pvals_bootstrap}
\centering
\resizebox{\textwidth}{!}{
\begin{tabular}{llcccc}
\toprule
\textbf{Model} & \textbf{Strategy} & \textbf{$\Delta$ Solve Rate (pp) [lo, hi]} & \textbf{p} & \textbf{$\Delta$ Mean Cost (\$) [lo, hi]} & \textbf{p} \\
\midrule
\multirow{2}{*}{Gemini 2.5 Flash} 
& Observation Masking  &  2.8 [-0.8, 6.4]  & 0.1504 & -0.2377 [-0.3202, -0.1614] & 0.0000\sig \\
& LLM-Summary          &  3.2 [-0.4, 7.0]  & 0.0948 & -0.1725 [-0.2579, -0.0936] & 0.0000\sig \\
\midrule
\multirow{2}{*}{\parbox{3.5cm}{Gemini 2.5 Flash\\(thinking)}} 
& Observation Masking  & -4.0 [-7.8, -0.2]  & 0.0406\sig & -0.3143 [-0.4096, -0.2245] & 0.0000\sig \\
& LLM-Summary          & -9.0 [-13.0, -5.2] & 0.0000\sig & -0.3046 [-0.4074, -0.2043] & 0.0000\sig \\
\midrule
\multirow{2}{*}{Qwen3-Coder 480B} 
& Observation Masking  &  1.4 [-1.6, 4.4]   & 0.3856 & -0.6762 [-0.9320, -0.4518] & 0.0000\sig \\
& LLM-Summary          &  0.4 [-3.0, 3.8]   & 0.8736 & -0.6491 [-0.9048, -0.4263] & 0.0000\sig \\
\midrule
\multirow{2}{*}{Qwen3-32B} 
& Observation Masking  & -2.0 [-5.0, 1.0]   & 0.2086 & -0.5632 [-0.7479, -0.3817] & 0.0000\sig \\
& LLM-Summary          & -1.0 [-4.6, 2.6]   & 0.6192 & -0.6174 [-0.7904, -0.4454] & 0.0000\sig \\
\midrule
\multirow{2}{*}{\parbox{3.5cm}{Qwen3-32B\\(thinking)}} 
& Observation Masking  &  1.6 [-2.0, 5.2]   & 0.3980 & -0.0510 [-0.1255,  0.0187] & 0.1586 \\
& LLM-Summary          &  1.8 [-1.8, 5.4]   & 0.3420 & -0.0021 [-0.0785,  0.0741] & 0.9370 \\
\bottomrule
\end{tabular}
}
\end{table}

\section{Additional Studies}
\label{appendix:ablations}
\begin{figure}[t]
    \centering
    \includegraphics[width=\linewidth]{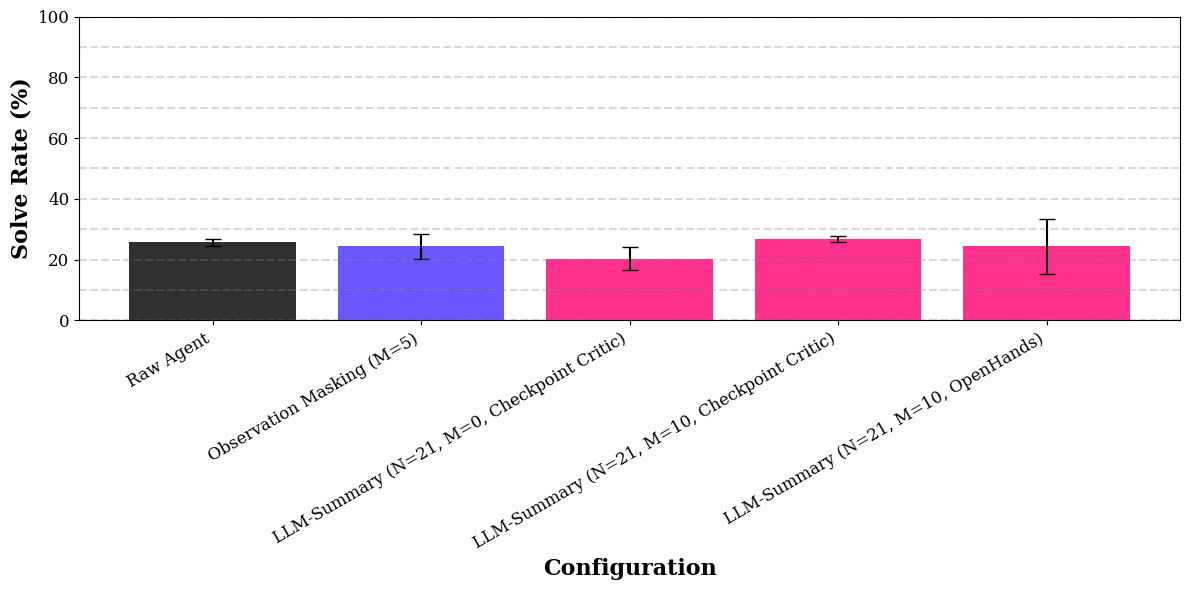}
    \caption{Downstream task performance a randomly generated 150-sample subset of SWE-bench Verified~\cite{openai_swe_bench_verified_2024} comparing the prompt in \Cref{appendix:llm-summary-prompt} with the joint critic-summarization prompt presented in this section. We find that simply prompting the model to include feedback in its summaries does not improve the solve rate and further increases the cost.}
    \label{fig:ablation-critic}
\end{figure}

For the critic-enhanced summarizer in \Cref{fig:ablation-critic}, we experiment on SWE-bench Lite-50~\cite{badertdinov2024scaling}. For the sensitivity to the rolling window size $M$ of the Observation Masking strategy and the configurations of the \ac{llm}-Summary strategy, we show our results on a randomly sampled 150-instance subset of SWE-bench Verified~\cite{openai_swe_bench_verified_2024} that we release with our code. We conduct these studies with GPT-4.1-mini~\cite{openai_gpt-4_2024}.

\begin{figure}[t]
    \centering
    \includegraphics[width=\linewidth]{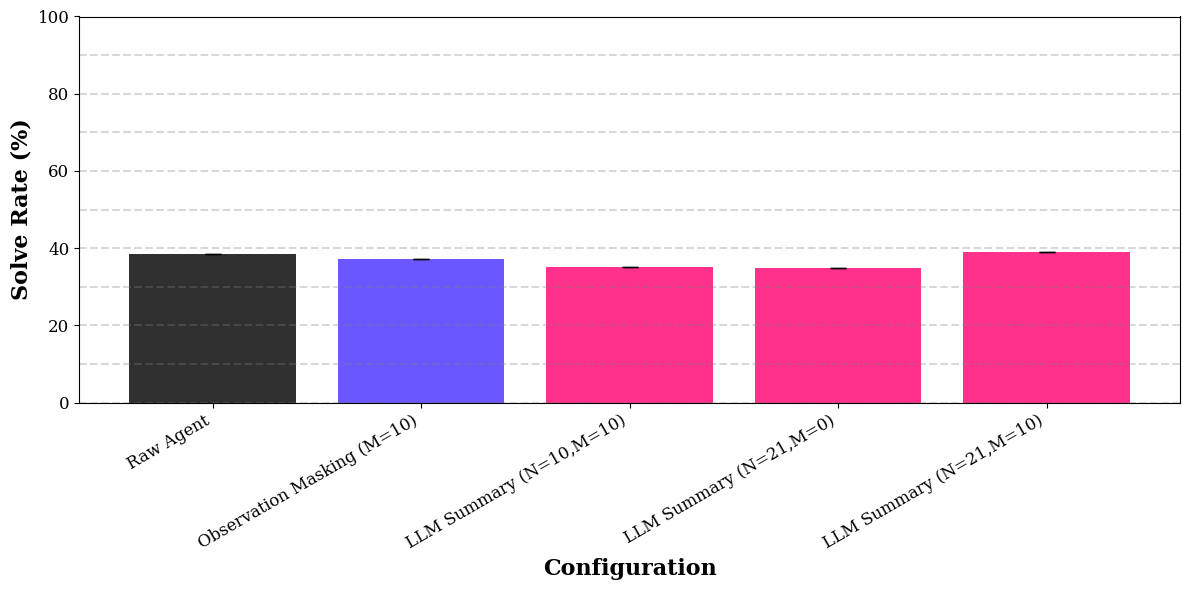}
    \caption{Downstream task performance of a single experiment on a randomly generated 150-sample subset of SWE-bench Verified~\cite{openai_swe_bench_verified_2024} across various different configuration combinations with respect to the tail length $M$. We find that a larger summarization window compared to the tail length improves performance.}
    \label{fig:ablation-tail-length}
\end{figure}

\subsection{Observation Masking Configuration}
\label{appendix:ablation:window-size}
\begin{figure}[t]
    \centering
    \includegraphics[width=\linewidth]{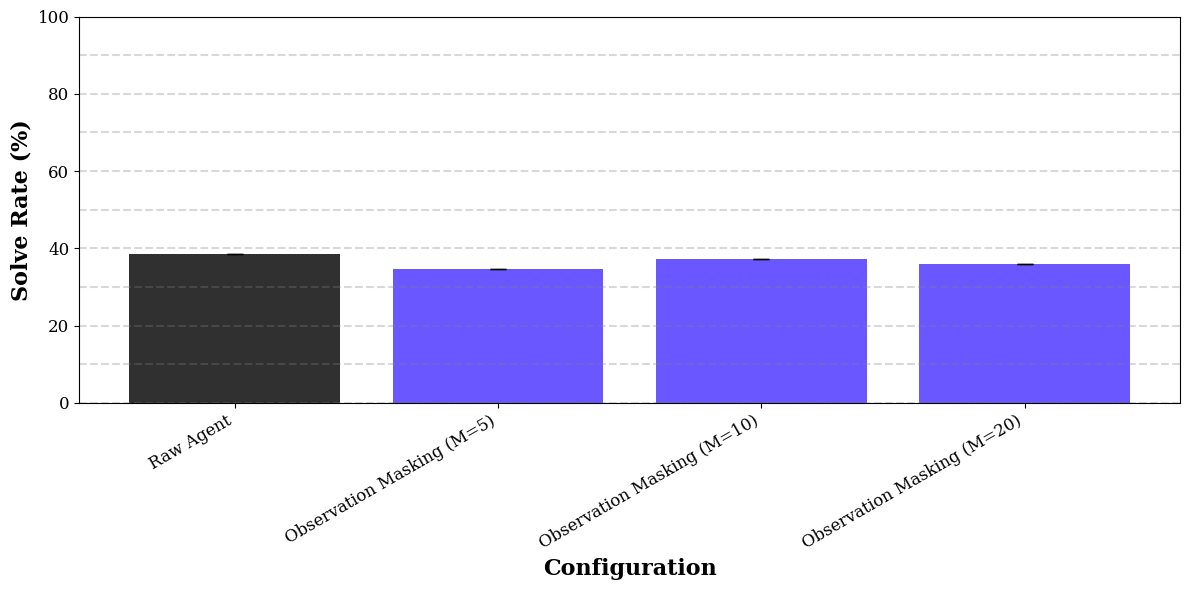}
    \caption{Downstream task performance of a single experiment on a randomly generated 150-sample subset of SWE-bench Verified~\cite{openai_swe_bench_verified_2024} across different context window sizes. We find that $M=10$ yields optimal performance.}
    \label{fig:ablation-window-size}
\end{figure}

We experiment with the rolling window size $M$ of the Observation Masking strategy. In \Cref{fig:ablation-window-size} we can see that the performance of the strategy peaks at $M=10$, before falling again when further increasing the window size to $M=20$. Thus, we use this configuration of the Observation Masking strategy for our main experiments.

\subsection{\ac{llm}-Summary Configuration}
\label{appendix:ablation:tail-length}
In \Cref{fig:ablation-tail-length} we show the solve rate of different experimental configurations for \ac{llm}-Summary in addition to those of our baselines. We find that using tail turns $M>0$ improves downstream performance. Furthermore, in contrast to the 50-50 split between turns to summarize and tail turns that OpenHands~\cite{wang2025openhands} uses, we find that summarizing more turns at once improves the solve rate. We thus proceed with $N=21, M=10$ for our main experiments.

% Main paper - discussion section
\subsection{Critic-Enhanced LLM-Summary} 
\label{appendix:ablation:critic}
A natural follow-up question is whether the \ac{llm}-Summary strategy could be improved by making the summarization process more intelligent. We explore this by enhancing our \ac{llm}-Summary strategy with execution-free feedback, a technique that has shown promise in scaling test-time compute for \ac{se} agents~\cite{pan_swegym_2024,jain_r2e-gym,zhang2025diversity,antoniades2025swesearch}. In this approach, the \ac{llm} simultaneously generates a summary and critical analysis of the trajectory, incorporating both into the compressed context. This is akin to providing reflections within a single rollout, instead of across multiple rollouts~\cite{shinn_reflexion_2023}.

In comparison to the modified OpenHands prompt in \Cref{fig:prompt:llm-summary}, we frame the task as generating a checkpoint instead of a summary and prompt the model to reflect on the turns to summarize. In doing so, we aim to encourage the model to generate an output that helps the agent adjust its solution path during an attempt and avoid overly grounding it in previous, potentially suboptimal or even flawed, turns through a plain summary.

To elicit meaningful reflections, we prompt the \ac{llm} with guiding questions that it could reflect and provide insights on. These questions assess whether the agent is stuck or looping, aligned with the initial problem statement, reflect on the agent's high-level solution approach with respect to the turns to summarize. Additionally, we provide few-shot examples to further guide the agents toward generating meaningful and actionable reflections~\cite{brown_language_2020}. Finally, as we did in the OpenHands-style prompt (\Cref{fig:prompt:llm-summary}), we provide the previous summary, or problem statement if none is available, and the turns to summarize to the model.

Testing on 150 samples from SWE-bench Verified~\cite{openai_swe_bench_verified_2024} using SWE-agent~\cite{yang_swe-agent_2024}, this critic-enhanced approach using the prompt presented in \Cref{fig:prompt:llm-critic-1,fig:prompt:llm-critic-2,fig:prompt:llm-critic-3} showed no improvement in solve rate over standard \ac{llm}-Summary. More concerning, we observed exacerbated trajectory elongation patterns, with critic-enhanced runs producing even longer trajectories than standard summarization. This is perhaps unsurprising: the critic's reflections naturally encourage the agent to explore alternative solution paths, try additional debugging strategies, or reconsider its approach, all of which translate to more turns, thus driving cost and reducing efficiency gains.

This finding reinforces our central insight about trajectory elongation. While execution-free feedback aims to improve agent decision-making, it paradoxically increases computational costs by extending exploration. The critic's guidance, rather than helping the agent efficiently recognize dead ends, provides additional avenues to pursue, further delaying termination. Furthermore, this increased cost, does not lead to increased downstream performance. This suggests that effective memory systems for AI agents require fundamental rethinking: simply adding more sophisticated feedback to summaries may compound rather than solve the efficiency challenges we identify.

\subsection{Behavior of the Covered Context Management Strategies Across Turns}
\label{appendix:preliminary-experiments}
\begin{figure}[th]
    \centering
    \includegraphics[width=\linewidth]{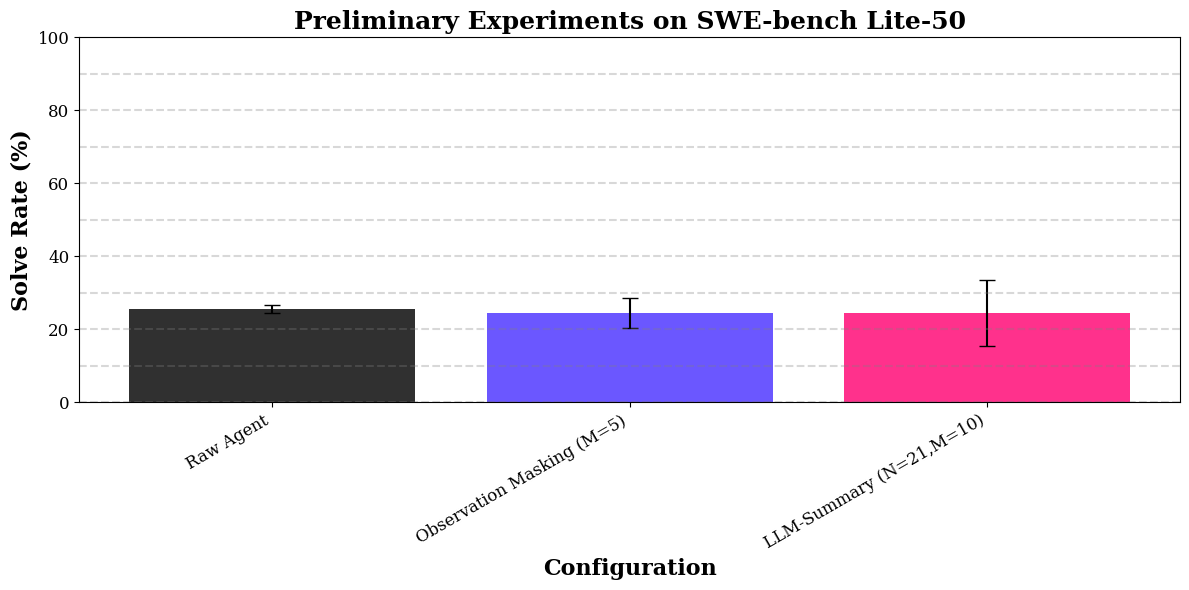}
    \caption{Downstream task performance on SWE-bench Lite-50\cite{badertdinov2024scaling} across the context management strategies we cover in our study. Bars represent the mean across three experiments, the error bars show the standard deviations. Surprisingly, $f_{RW}$ performs on par with the \ac{llm}-based strategy using more compute.}
    \label{fig:preliminary_experiments-solve-rate}
\end{figure}

In \Cref{fig:preliminary_experiments-solve-rate}, we show preliminary experimental results using the trajectory management strategies introduced in Section \ref{subsec:trajectory-management-strategies} on SWE-bench Lite-50~\cite{badertdinov2024scaling} with GPT-4.1-mini~\cite{gpt_41_tech_report}. Here, we use a rolling window size of $M=5$, following SWE-agent~\cite{yang_swe-agent_2024} and $N=21,M=10$ for the \ac{llm}-based approach paired with a slightly modified version of OpenHand's prompt~\cite{wang2025openhands} (see Appendix~\ref{appendix:llm-summary-prompt}). We observe that the Observation Masking strategy poses a strong baseline, consistently performing equally or better on the downstream task on SWE-bench Lite-50~\cite{badertdinov2024scaling} than the \ac{llm}-Summary approach despite using being much simpler. 

\begin{figure}[th]
    \centering
    \includegraphics[width=\linewidth]{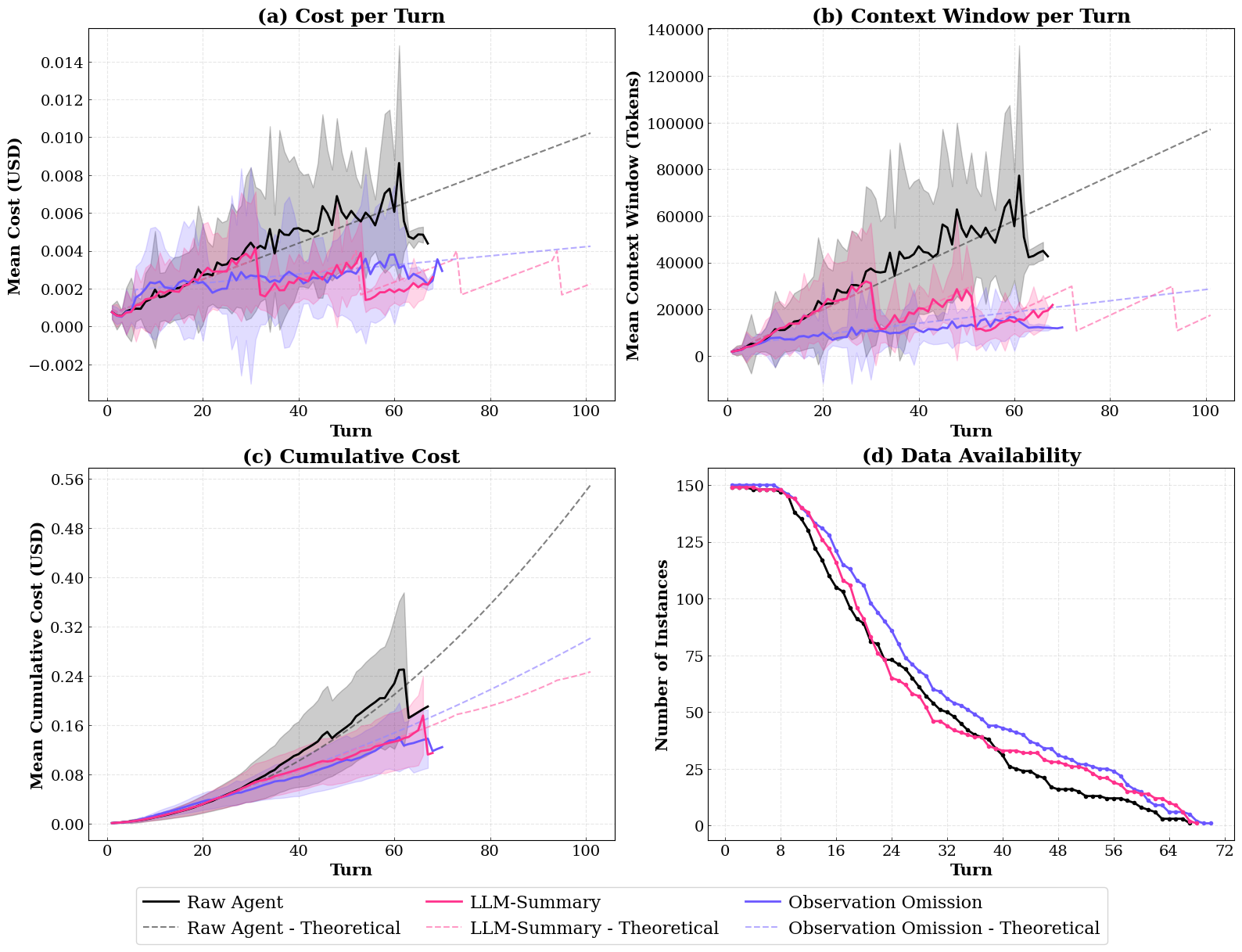}
    \caption{\textbf{Preliminary experimental results motivating our study.} Dashed lines show expected behavior based on mean token counts per turn type observed in the raw agent trajectories. We micro-average all results with standard deviation shown as shaded regions. \textbf{(a),(b), and (c)} The observed data closely match the simulated effects of applying either trajectory management strategy to the simulated raw agent trajectory. The effects of the \ac{llm} and Observation Masking approach on cost and context window size overlap at lower turn numbers. Due to the bounding of the context-window we expect the \ac{llm} approach to be especially effective on long trajectories. \textbf{(d)} With an increasing number of turns, our empiric data becomes increasingly sparse.}
    \label{fig:preliminary_experiments}
\end{figure}

To investigate why this is the case, and uncover potential scenarios in which the \ac{llm}-Summary strategy may be beneficial, we analyze the behavior of these strategies across turns in \Cref{fig:preliminary_experiments}. The solid colored lines are the empirically observed results using SWE-agent~\cite{yang_swe-agent_2024}. For each trajectory management strategy, we run three experiments, yielding 150 trajectories total. To visualize these data, we use the micro-averaged mean for each turn. The dashed lines indicate the empirically grounded simulated behavior. To generate the data for these simulated trajectories, we compute the mean token consumption per token type across all experimental data available for the raw agent:

\begin{equation}
\label{eq:simulated-traj}
\bar{x} = \frac{1}{T_{total}} \sum_{i=1}^{3}\sum_{j=1}^{50}\sum_{k=1}^{T_{local}}x_{ijk} \quad \text{where} \quad
x \in \{r,a,o\}
\end{equation}

where $T_{total}$ is the total number of turns $T$ we observed across all instances and experiments and $T_{local}$ is the number of turns of a single trajectory. We then generate a turn $T_{sim}=(\bar{r}, \bar{a}, \bar{o})$ using placeholder tokens. By repeatedly appending $T_{sim}$ we generate a simulated agent trajectory $\tau_{sim}=(T_{sim},\ldots,T_{sim})$ of arbitrary length. To generate the data for the simulated \ac{llm}-Summary and Observation Masking context management trajectories, we apply these strategies to $\tau_{sim}$. This allows us to study the expected behavior of these trajectory management strategies up to a large number of turns.

Figures \ref{fig:preliminary_experiments} a, b and c show that our experimental data match the expected behaviour of simulated trends closely. Surprisingly, we find that the \textbf{$f_{RW}$ is competitive with the the \ac{llm}-based strategy on cost and even outperforms it on context compression}.

$f_{RW}$ is a strong baseline, due to the distribution of tokens across the types $r,a,o$. In~\Cref{fig:token-type-distr} we plot the share of token types in $T_{sim}$. The environment observation tokens $o$ overwhelmingly dominate the composition of $T_{sim}$, contributing $\approx84\%$. Thus targeting this token type is extremely effective. 

We can see this in effect in Figures \ref{fig:preliminary_experiments}a, and b. While the cost of the two context management strategies is similar, due to the worse cache behavior of $f_{RW}$, $f_{RW}$ offers superior compression especially at lower turn numbers. Looking at our simulations on the other hand, we expect the \ac{llm}-based approach to start outperforming $f_{RW}$ on longer trajectories because it bounds the maximum context size in a fuzzy manner, resulting in a saw-function for both the cost and context window size. This motivates us to set the turn limit in our main experiments to 250.

\section{\ac{llm} Summary Prompts}
\label{appendix:llm-summary-prompt}
We share our prompt template for summary generation in \Cref{fig:prompt:llm-summary}. Compared to OpenHands~\cite{wang2025openhands}, we remove the part of the prompt that aims to handle summary generation for tasks outside the \ac{se} domain, since our work is purely focused on the \ac{se} domain. In addition to the system prompt shown in \Cref{fig:prompt:llm-summary}, we provide a joint critic-summarization prompt in \Cref{fig:prompt:llm-critic-1,fig:prompt:llm-critic-2,fig:prompt:llm-critic-3}. We discuss the effects of generating execution-free in \Cref{appendix:ablation:critic} and \Cref{sec:discussion}.

\begin{figure*}[th]
\begin{tcolorbox}[
  colback=jb_magenta_accent,
  colframe=jb_magenta,
  title=\ac{llm}-Summary Prompt, width=\textwidth
]
\begin{small}
\begin{verbatim}
You are maintaining a context-aware state summary for an interactive agent. 
You will be given a list of events corresponding to actions taken by the 
agent, and the most recent previous summary if one exists. Track:

USER_CONTEXT: (Preserve essential user requirements, goals, and 
clarifications in concise form)

COMPLETED: (Tasks completed so far, with brief results)
PENDING: (Tasks that still need to be done)
CURRENT_STATE: (Current variables, data structures, or relevant state)

For code-specific tasks, also include:
CODE_STATE: (File paths, function signatures, data structures)
TESTS: (Failing cases, error messages, outputs)
CHANGES: (Code edits, variable updates)
DEPS: (Dependencies, imports, external calls)
VERSION_CONTROL_STATUS: (Repository state, current branch, PR status, 
commit history)

PRIORITIZE:
1. Adapt tracking format to match the actual task type
2. Capture key user requirements and goals
3. Distinguish between completed and pending tasks
4. Keep all sections concise and relevant

SKIP: Tracking irrelevant details for the current task type

Example formats:

For code tasks:
USER_CONTEXT: Fix FITS card float representation issue
COMPLETED: Modified mod_float() in card.py, all tests passing
PENDING: Create PR, update documentation
CODE_STATE: mod_float() in card.py updated
TESTS: test_format() passed
CHANGES: str(val) replaces f"{val:.16G}"
DEPS: None modified
VERSION_CONTROL_STATUS: Branch: fix-float-precision, Latest commit: a1b2c3d

<PREVIOUS_SUMMARY>
...
</PREVIOUS_SUMMARY>
<TURN-0>
...
</TURN-0>
...
<TURN-20>
...
</TURN-20>
\end{verbatim}
\end{small}
\end{tcolorbox}
\caption{The \ac{llm}-Summary prompt we use in SWE-agent~\cite{yang_swe-agent_2024} is a slightly modified version of the OpenHands \ac{llm}-Summary system prompt~\cite{wang2025openhands}. Additionally we pass the previous summary and the turns to summarize. If no previous summary is available we instead pass the task problem statement as initial context for the summary generation.}
\label{fig:prompt:llm-summary}
\end{figure*}

\begin{figure*}[th]
\begin{tcolorbox}[
  colback=jb_magenta_accent,
  colframe=jb_magenta,
  title=Joint Critic-Summary Prompt (Part 1), width=\textwidth
]
\begin{small}
\begin{verbatim}
You are maintaining a context-aware state checkpoint for an interactive agent
working on software engineering tasks (specifically, bug fixes), assessing the
agents progress toward completing the task, and offering suggestions and 
guidance if it is not on track. 
You will be given a list of turns corresponding to actions taken by the agent
and their resulting observations, and the most recent previous checkpoint if
one exists. You must proceed in the following two phases:
1. <CHECKPOINT>
2. <REFLECTIONS>

A <CHECKPOINT> should capture the current repository state and the agent's 
progress towards completing the task. It consists of:
USER_CONTEXT: (Preserve essential user requirements, goals, and clarifications
based on findings, previous checkpoints, and the initial problem statement in 
concise form)
CODE_STATE: (File paths, function signatures, data structures followed by 
their current state)
TESTS: (Failing cases, error messages, outputs)
CHANGES: (Code edits, variable updates)
DEPS: (Dependencies, imports, external calls)

PRIORITIZE:
1. Capture key requirements from the initial issue description or previous 
checkpoints and reflections
2. Keep all sections concise and relevant to fixing the issue
3. Focus on information that quantifies the agent's progress towards a solution

Next you must reflect on the agent's progres in the below to turns to 
generate a set of <REFLECTIONS>. Here are some aspects to consider 
when generating the <REFLECTIONS>:
- Are the agent's actions still aligned with the initial user requirement?
- Is the agent making progress?
- Is it stuck in a loop or repeatedly carrying out the same actions?
- Can you identify any problematic patterns in the agent's actions?
- Did the agent follow the issue description or your previous feedback? 
    If so, why did or didn't it make meaningful progress?
- Which critical piece of information might help the agent get back on 
    track?
- What has the agent not tried so far?

Key requirements for the <REFLECTIONS>: 
1. Avoid reporting nitpicks as they may confuse the agent.
2. Your reflections should be diverse with respect to any available previous
reflections.
3. Provide up to 2 reflections total. Reflections are mutually exclusive.
4. Each reflection should identify one distinct problem and may include one
or two fixes for that problem.
5. Limit yourself to the most critical issues that are blocking progress.

When generating the <REFLECTIONS>, follow the format below:
<REFLECTIONS>
Problem-A: (a detailed description of a problem the agent is facing)
Fix-A.1: (proposed solution, guidance or hint for overcoming the problem 
including the rationale for it)
</REFLECTIONS>
\end{verbatim}
\end{small}
\end{tcolorbox}
\caption{Part 1 of our joint critic and summarization \ac{llm}-Summary prompt. Compared to the \ac{llm}-Summary prompt we use in our main experiments (\Cref{fig:prompt:llm-summary}), we also prompt the \ac{llm} to act as execution-free critic regarding the turns it is summarizing.}
\label{fig:prompt:llm-critic-1}
\end{figure*}

\begin{figure*}[th]
\begin{tcolorbox}[
  colback=jb_magenta_accent,
  colframe=jb_magenta,
  title=Joint Critic-Summary Prompt (Part 2), width=\textwidth
]
\begin{small}
\begin{verbatim}
Example output and format:
<CHECKPOINT>
USER_CONTEXT: Fix failing authentication in REST API. Users report 
"Invalid token" errors after ~30 minutes of activity. The API should 
maintain user sessions properly with JWT tokens that expire after 1 hour.
CODE_STATE: 
1. api/auth_middleware.py: validate_token() MODIFIED with logging.
2. api/auth_utils.py: refresh_token() MODIFIED to auto-refresh at 45 min.
3. config.py: JWT_EXPIRATION unchanged at 3600.
TESTS:
1. tests/test_auth_integration.py::test_long_session: FAILING - "Token 
expired at 32 minutes"
2. tests/test_auth_integration.py::test_token_refresh: PASSING
3. tests/test_auth_unit.py: ALL PASSING
CHANGES:
1. auth_utils.py: Added auto-refresh logic when token age > 2700 seconds.
2. auth_middleware.py: Added debug logging for token validation steps.
DEPS: PyJWT==2.4.0, python-jose==3.3.0 (both imported)
</CHECKPOINT>
<REFLECTIONS>
Problem-A: Agent has spent 6 turns modifying the token refresh logic 
and adding complex auto-refresh mechanisms, but hasn't investigated why
tokens are expiring at ~30 minutes when they're configured for 60 minutes.
The agent is treating the symptom (early expiration) by adding refresh 
logic, rather than finding the root cause. The fact that tokens consistently 
expire at 30-32 minutes suggests either: (1) a configuration mismatch 
somewhere else overriding the 3600-second setting, (2) a timezone/clock 
issue between server and client, or (3) the JWT library might be using a 
different time unit or has a default max age.
Fix-A.1: Stop adding refresh logic and investigate the actual token 
expiration time. Add logging to print the exact 'exp' claim value when 
tokens are created and when they're validated. Check if there's another
config file, environment variable, or hardcoded value setting token 
expiration to 1800 seconds (30 min). Also verify the JWT library's time unit 
- some libraries use milliseconds while others use seconds. The issue is 
likely a simple configuration problem, not a need for complex refresh 
mechanisms.
Fix-A.2: The solution the agent is trying to implement is overly complex, 
confusing, and not tackeling the root cause. The agent should take a step 
back and think about what it is actually trying to do, discard its current
approach and come up with a simpler solution that is more likely to work. 
It should recall SE best practices and clean code principles.

Problem-B: Agent has successfully modified token validation and refresh logic, 
but the integration test continues to fail at exactly 32 minutes. Despite the
previous guidance to investigate configuration mismatches, the agent hasn't 
checked for environmental differences between unit tests (which pass) and 
integration tests (which fail). The agent also hasn't noticed that two
different JWT libraries are imported (PyJWT and python-jose), which could
mean tokens are created with one library but validated with another. 
Additionally, the agent keeps focusing on server-side fixes without considering
that the test client might have its own timeout or token handling logic 
that's causing the consistent 32-minute failure.
Fix-B.1: Audit which JWT library is actually being used where. Search for 
`from jose import` and `from jwt import` patterns across the codebase. Create
a simple debug endpoint that generates a token and immediately decodes it 
with both libraries to see if they interpret expiration differently. The 
symptom of tokens expiring at ~32 minutes (close to but not exactly 30) 
could indicate timestamp precision or timezone handling differences between 
the libraries. Consider standardizing on one JWT library throughout the codebase.
</REFLECTIONS>
\end{verbatim}
\end{small}
\end{tcolorbox}
\caption{Part 2 of our joint critic and summarization \ac{llm}-Summary prompt. Compared to the \ac{llm}-Summary prompt we use in our main experiments (\Cref{fig:prompt:llm-summary}), we also prompt the \ac{llm} to act as execution-free critic regarding the turns it is summarizing.}
\label{fig:prompt:llm-critic-2}
\end{figure*}

\begin{figure*}[th]
\begin{tcolorbox}[
  colback=jb_magenta_accent,
  colframe=jb_magenta,
  title=Joint Critic-Summary Prompt (Part 3), width=\textwidth
]
\begin{small}
\begin{verbatim}
<PREVIOUS_SUMMARY>
...
</PREVIOUS_SUMMARY>
<TURN-0>
...
</TURN-0>
...
<TURN-20>
...
</TURN-20>
\end{verbatim}
\end{small}
\end{tcolorbox}
\caption{Part 3 of our joint critic and summarization \ac{llm}-Summary prompt. Compared to the \ac{llm}-Summary prompt we use in our main experiments (\Cref{fig:prompt:llm-summary}), we also prompt the \ac{llm} to act as execution-free critic regarding the turns it is summarizing.}
\label{fig:prompt:llm-critic-3}
\end{figure*}

%%%%%%%%%%%%%%%%%%%%%%%%%%%%%%%%%%%%%%%%%%%%%%%%%%%%%%%%%%%%

\clearpage
\section*{NeurIPS Paper Checklist}

\begin{enumerate}

\item {\bf Claims}
    \item[] Question: Do the main claims made in the abstract and introduction accurately reflect the paper's contributions and scope?
    \item[] Answer: \answerYes{} % Replace by \answerYes{}, \answerNo{}, or \answerNA{}.
    \item[] Justification: The claims made in the abstract and introduction are discussed in detail in \Cref{sec:main-results} and \Cref{sec:discussion}. Additionally, we extensively document our experimental setup in \Cref{sec:experimental-setup} and take steps to ensure our experiments cover diverse configurations and find that our experimental results support our claims across all configurations.

\item {\bf Limitations}
    \item[] Question: Does the paper discuss the limitations of the work performed by the authors?
    \item[] Answer: \answerYes{} % Replace by \answerYes{}, \answerNo{}, or \answerNA{}.
    \item[] Justification: We explicitly document the limitations of our work and the threats to the validity of our study in \Cref{sec:limitations}.

\item {\bf Theory assumptions and proofs}
    \item[] Question: For each theoretical result, does the paper provide the full set of assumptions and a complete (and correct) proof?
    \item[] Answer: \answerNA{} % Replace by \answerYes{}, \answerNo{}, or \answerNA{}.
    \item[] Justification: Our work presents an extensive, empiric study of the cost-performance tradeoff of \ac{se} agent context management and thus does not contain theoretical results.

    \item {\bf Experimental result reproducibility}
    \item[] Question: Does the paper fully disclose all the information needed to reproduce the main experimental results of the paper to the extent that it affects the main claims and/or conclusions of the paper (regardless of whether the code and data are provided or not)?
    \item[] Answer: \answerYes{} % Replace by \answerYes{}, \answerNo{}, or \answerNA{}.
    \item[] Justification: In \Cref{sec:experimental-setup} we comprehensively document our experimental setup, including the models used, their hyperparameters, and the benchmark we evaluate on. Furthermore, we provide details on our chosen configurations in \Cref{appendix:preliminary-experiments} and \Cref{appendix:ablations}.

\item {\bf Open access to data and code}
    \item[] Question: Does the paper provide open access to the data and code, with sufficient instructions to faithfully reproduce the main experimental results, as described in supplemental material?
    \item[] Answer: \answerYes{} % Replace by \answerYes{}, \answerNo{}, or \answerNA{}.
    \item[] Justification: We release our \ac{llm}-summary implementation\footnote{\url{https://github.com/JetBrains-Research/the-complexity-trap}} for SWE-agent~\cite{yang_swe-agent_2024}. The experimental results of our main experiments can be openly accessed via HuggingFace\footnote{\url{https://huggingface.co/datasets/JetBrains-Research/the-complexity-trap}}.

\item {\bf Experimental setting/details}
    \item[] Question: Does the paper specify all the training and test details (e.g., data splits, hyperparameters, how they were chosen, type of optimizer, etc.) necessary to understand the results?
    \item[] Answer: \answerYes{}{} % Replace by \answerYes{}, \answerNo{}, or \answerNA{}.
    \item[] Justification: In \Cref{sec:experimental-setup} we comprehensively document our experimental setup, including the models used, their hyperparameters, and the benchmark we evaluate on. Furthermore, we provide details on our chosen configurations in \Cref{appendix:preliminary-experiments} and \Cref{appendix:ablations}.

\item {\bf Experiment statistical significance}
    \item[] Question: Does the paper report error bars suitably and correctly defined or other appropriate information about the statistical significance of the experiments?
    % TODO: not sure if this will be enough, but I dont have time to bootstrap CIs or something like that for the main results right now I think.
    \item[] Answer: \answerYes{} % Replace by \answerYes{}, \answerNo{}, or \answerNA{}.
    \item[] Justification: In \Cref{sec:experimental-setup} we show the standard deviation of the solve rate, cost and context window size in our preliminary experiments.

\item {\bf Experiments compute resources}
    \item[] Question: For each experiment, does the paper provide sufficient information on the computer resources (type of compute workers, memory, time of execution) needed to reproduce the experiments?
    \item[] Answer: \answerYes{} % Replace by \answerYes{}, \answerNo{}, or \answerNA{}.
    \item[] Justification: We present details on our infrastructure in \Cref{sec:experimental-setup}.
    
\item {\bf Code of ethics}
    \item[] Question: Does the research conducted in the paper conform, in every respect, with the NeurIPS Code of Ethics \url{https://neurips.cc/public/EthicsGuidelines}?
    \item[] Answer: \answerYes{} % Replace by \answerYes{}, \answerNo{}, or \answerNA{}.
    \item[] Justification: Our findings pave the way toward effective and efficient \ac{llm} agents, offering an immediate way of reducing the impact of AI on our climate and environment.

\item {\bf Broader impacts}
    \item[] Question: Does the paper discuss both potential positive societal impacts and negative societal impacts of the work performed?
    \item[] Answer: \answerYes{} % Replace by \answerYes{}, \answerNo{}, or \answerNA{}.
    \item[] Justification: We highlight the positive impact of reduced computational costs on our environment in \Cref{sec:conclusion}.
    
\item {\bf Safeguards}
    \item[] Question: Does the paper describe safeguards that have been put in place for responsible release of data or models that have a high risk for misuse (e.g., pretrained language models, image generators, or scraped datasets)?
    \item[] Answer: \answerNA{}{} % Replace by \answerYes{}, \answerNo{}, or \answerNA{}.
    \item[] Justification: Our work presents an extensive evaluation regarding the efficiency of existing \ac{se} agent systems, using existing models and an existing benchmark and thus does not pose any such risk.

\item {\bf Licenses for existing assets}
    \item[] Question: Are the creators or original owners of assets (e.g., code, data, models), used in the paper, properly credited and are the license and terms of use explicitly mentioned and properly respected?
    \item[] Answer: \answerYes{} % Replace by \answerYes{}, \answerNo{}, or \answerNA{}.
    \item[] Justification: We credit the authors of SWE-bench~\cite{jimenez2024swebench}, SWE-bench Lite-50~\cite{badertdinov2024scaling}, SWE-bench Verified~\cite{openai_swe_bench_verified_2024}, SWE-agent~\cite{yang_swe-agent_2024} and OpenHands~\cite{wang2025openhands} at the appropriate locations in our work in \Cref{sec:related-work}, \Cref{sec:experimental-setup}, and \Cref{sec:main-results}.

\item {\bf New assets}
    \item[] Question: Are new assets introduced in the paper well documented and is the documentation provided alongside the assets?
    \item[] Answer: \answerYes{} % Replace by \answerYes{}, \answerNo{}, or \answerNA{}.
    \item[] Justification: We release our implementation for generating \ac{llm}-summaries in SWE-agent~\cite{yang_swe-agent_2024} with this work in a well-documented manner.

\item {\bf Crowdsourcing and research with human subjects}
    \item[] Question: For crowdsourcing experiments and research with human subjects, does the paper include the full text of instructions given to participants and screenshots, if applicable, as well as details about compensation (if any)? 
    \item[] Answer: \answerNA{} % Replace by \answerYes{}, \answerNo{}, or \answerNA{}.
    \item[] Justification: All of our experiments are conducted on existing benchmarks. Our work does not involve any crowdsourced labor.

\item {\bf Institutional review board (IRB) approvals or equivalent for research with human subjects}
    \item[] Question: Does the paper describe potential risks incurred by study participants, whether such risks were disclosed to the subjects, and whether Institutional Review Board (IRB) approvals (or an equivalent approval/review based on the requirements of your country or institution) were obtained?
    \item[] Answer: \answerNA{} % Replace by \answerYes{}, \answerNo{}, or \answerNA{}.
    \item[] Justification: All of our experiments are conducted on existing benchmarks. Our work does not involve any crowdsourced labor.

\item {\bf Declaration of LLM usage}
    \item[] Question: Does the paper describe the usage of LLMs if it is an important, original, or non-standard component of the core methods in this research? Note that if the LLM is used only for writing, editing, or formatting purposes and does not impact the core methodology, scientific rigorousness, or originality of the research, declaration is not required.
    %this research? 
    \item[] Answer: \answerYes{} % Replace by \answerYes{}, \answerNo{}, or \answerNA{}.
    \item[] Justification: Our work focuses on the efficiency on AI agents in \ac{se}, thus we explicitly discuss the use of specific \acp{llm} throughout our work, for example in \Cref{sec:experimental-setup} and \Cref{sec:main-results}.

\end{enumerate}

\end{document}